\documentclass[lettersize,journal]{IEEEtran}
\usepackage{amsmath,amsfonts}
\usepackage{algorithm}
\usepackage{array}
\usepackage[caption=false,font=normalsize,labelfont=sf,textfont=sf]{subfig}
\usepackage{textcomp}
\usepackage{stfloats}
\usepackage{url}
\usepackage{verbatim}
\usepackage{graphicx}
\usepackage{cite}
\usepackage{balance}

\newcommand{\para}[1]{\vspace{2pt}\noindent\textbf{#1.~}}
\newcommand{\ignore}[1]{}
\newcommand{\system}{\sloppy{SmartHalo\@}}

\usepackage{ulem}
\usepackage[switch]{lineno}
\newcommand{\publicUrl}{\url{https://github.com/Janelinux/SmartHalo}}
\usepackage{listings, xcolor}
\definecolor{verylightgray}{rgb}{.97,.97,.97}
\lstdefinelanguage{Solidity}{
  keywords=[1]{anonymous, assembly, assert, balance, break, call, callcode, case, catch, class, constant, continue, constructor, contract, debugger, default, delegatecall, delete, do, else, emit, event, experimental, export, external, false, finally, for, function, gas, if, implements, import, in, indexed, instanceof, interface, internal, is, length, library, log0, log1, log2, log3, log4, memory, modifier, new, payable, pragma, private, protected, public, pure, push, require, return, returns, revert, selfdestruct, send, solidity, storage, struct, suicide, super, switch, then, this, throw, transfer, true, try, typeof, using, value, view, while, with, addmod, ecrecover, keccak256, mulmod, ripemd160, sha256, sha3}, 
  keywordstyle=[1]\color{blue}\bfseries,
  keywords=[2]{address, bool, byte, bytes, bytes1, bytes2, bytes3, bytes4, bytes5, bytes6, bytes7, bytes8, bytes9, bytes10, bytes11, bytes12, bytes13, bytes14, bytes15, bytes16, bytes17, bytes18, bytes19, bytes20, bytes21, bytes22, bytes23, bytes24, bytes25, bytes26, bytes27, bytes28, bytes29, bytes30, bytes31, bytes32, enum, int, int8, int16, int24, int32, int40, int48, int56, int64, int72, int80, int88, int96, int104, int112, int120, int128, int136, int144, int152, int160, int168, int176, int184, int192, int200, int208, int216, int224, int232, int240, int248, int256, mapping, string, uint, uint8, uint16, uint24, uint32, uint40, uint48, uint56, uint64, uint72, uint80, uint88, uint96, uint104, uint112, uint120, uint128, uint136, uint144, uint152, uint160, uint168, uint176, uint184, uint192, uint200, uint208, uint216, uint224, uint232, uint240, uint248, uint256, var, void, ether, finney, szabo, wei, days, hours, minutes, seconds, weeks, years},  
  keywordstyle=[2]\color{teal}\bfseries,
  keywords=[3]{block, blockhash, coinbase, difficulty, gaslimit, number, timestamp, msg, data, gas, sender, sig, value, now, tx, gasprice, origin},  
  keywordstyle=[3]\color{violet}\bfseries,
  identifierstyle=\color{black},
  sensitive=false,
  comment=[l]{//},
  morecomment=[s]{/*}{*/},
  commentstyle=\color{gray}\ttfamily,
  stringstyle=\color{red}\ttfamily,
  morestring=[b]',
  morestring=[b]"
}
\usepackage{pifont}
\usepackage{oplotsymbl}
\usepackage{siunitx}
\usepackage{array,framed}
 \usepackage{
   color,
   float,
   epsfig,
   wrapfig,
   graphics,
   graphicx,
   subcaption
 }
\usepackage{setspace}
\usepackage{amsfonts}
\usepackage{latexsym,fancyhdr,url}
\usepackage{enumerate}
\usepackage{algpseudocode}
\usepackage{graphics}
\usepackage{xparse} 
\usepackage{xspace}
\usepackage{multirow}
\usepackage{microtype}

\usepackage{fancyhdr}

\begin{document}

\title{Augmenting Smart Contract Decompiler Output through Fine-grained Dependency Analysis and LLM-facilitated Semantic Recovery}

\author{
Zeqin Liao, 
Yuhong Nan,~\IEEEmembership{Member,~IEEE}, 
Zixu Gao, 
Henglong Liang, 
Sicheng Hao, 
Peifan Ren, 
Zibin Zheng,~\IEEEmembership{Fellow,~IEEE}         
\thanks{ 
\setlength{\parindent}{0pt}
Zeqin~Liao, Yuhong~Nan, Zixu~Gao, Henglong~Liang, Sicheng~Hao, Peifan~Ren and~Zibin~Zheng are with School of Software Engineering, Sun Yat-sen University, China and GuangDong Engineering Technology Research Center
of Blockchain, China. 

E-mail: \{liaozq8, gaozx9, lianghlong, haosch, renpf\} @mail2.sysu.edu.cn

E-mail: \{nanyh, zhzibin\}@mail.sysu.edu.cn } 

\thanks{
\setlength{\parindent}{0pt}
Yuhong Nan is the corresponding author}

}

\markboth{Journal of \LaTeX\ Class Files,~Vol.~14, No.~8, August~2021}%
{Shell \MakeLowercase{\textit{et al.}}: A Sample Article Using IEEEtran.cls for IEEE Journals}


\maketitle

\begin{abstract}
Decompiler is a specialized type of reverse engineering tool extensively employed in program analysis tasks,
particularly in program comprehension and vulnerability detection. However, current Solidity smart contract decompilers face significant limitations in reconstructing the original source code. In particular, the bottleneck of SOTA decompilers lies in inaccurate function identification, incorrect variable type recovery, and missing contract attributes. These deficiencies hinder downstream tasks and understanding of the program logic. To address these challenges, we propose SmartHalo, a new framework that enhances decompiler output by combining static analysis (SA) and large language models (LLM). SmartHalo leverages the complementary strengths of SA's accuracy in control and data flow analysis and LLM's capability in semantic prediction. More specifically, \system{} constructs a new data structure - Dependency Graph (DG), to extract semantic dependencies via static analysis. Then, it takes DG to create prompts for LLM optimization. Finally, the correctness of LLM outputs is validated through symbolic execution and formal verification.
Evaluation on a dataset consisting of 465 randomly selected smart contract functions shows that SmartHalo significantly improves the quality of the decompiled code, compared to SOTA decompilers (e.g., Gigahorse). Notably, integrating GPT-4o mini with SmartHalo further enhances its performance,  achieving a precision of 91.32\% and a recall of 87.38\% for function boundaries, a precision of 90.40\% and a recall of 88.82\% for variable types, and a precision of 80.66\% and a recall of 91.78\% for contract attributes.
\end{abstract}


\section{Introduction}
\label{sec:intro}



Decompiler is a specific type of reverse engineering tool widely used for program analysis, such as program comprehension and vulnerability detection~\cite{liao2024smartaxe}. For smart contracts written in Solidity, the goal of the decompiler is to recover the machine-executable code (e.g., the EVM bytecode~\cite{EVM}) back to the original source code written by developers (e.g., the Solidity program~\cite{Solidity}).

While decompilers recover abstractions that enhance
code readability and are widely used by reverse engineers, decompilers are incapable of fully reconstructing the original developer-written code, as the compiler irreversibly subverts crucial information for optimization~\cite{hu2024degpt}. In particular, critical information such as variable types, function boundaries, variable names, and annotation are absent in contract bytecode, and are unrecoverable by decompilers. 
As pointed out by prior research~\cite{liao2022SmartDagger,li2024varlifter,yang2024uncover, liao2025satellite}, these limitations make the
state-of-the-art solidity decompilers, such as Gigahorse~\cite{grech2019Gigahorse} and Etherscan decompiler~\cite{etherscan}, share the following three deficiency in their output: (1) inaccurate function identification: current decompilers often fail to precisely determine function boundaries, leading to the omission of important functions or erroneous function range identification. For example, decompilers often incorrectly recover multiple functions as a single function, which obviously increases complexity in certain tasks (e.g., function-level similarity comparison). (2) inaccurate variable type recovery: decompilers may produce type errors that are inconsistent with static domain rules. For instance, the decompiler often ignores predefined type (e.g., the return value of hash function \textit{keccak256} is the type of \textit{bytes32}), while recovering them as the type of \textit{uint256} uniformly.  (3) lack of contract attributes. Smart contracts employ state variables to record critical contract attributes (e.g., \textit{Asset}, \textit{identity}, \textit{router}). Although these contract attributes are explicitly stated in the source code through meaningful variable names and annotations, they are entirely omitted in the decompiled code. As highlighted in prior work~\cite{liao2022SmartDagger}, contract attribute is crucial for vulnerability detection tasks. 




Recent advancements~\cite{zhao2023DeepInfer,chen2022SigRec,li2024varlifter,liao2022SmartDagger,he2023neural} start optimizing or reconstructing the absent contract information such as variable type by analyzing the context of decompiler output,  even when these information is not part of the contract bytecode.
%
%
These works are far from satisfied, due to their limited scope and approach-wise weaknesses. 
More specifically, SmartDagger~\cite{liao2022SmartDagger}, a framework for detecting cross-contract vulnerability, can partially recover contract attributes (e.g., asset) for state variables from the decompiled bytecode. 
Neural-FEBI~\cite{he2023neural} is designed for identifying function boundaries, but it does not support boundary recovery for complex functions such as modifier functions or functions inherited from other contracts/subcontracts.
SigRec~\cite{chen2022SigRec}, VarLifter~\cite{li2024varlifter} and DeepInfer~\cite{zhao2023DeepInfer} can partially recover the variable types such as parameter types of already-known function signatures. However, SigRec~\cite{chen2022SigRec} and VarLifter~\cite{li2024varlifter} are heavily relied on pre-defined heuristics according to EVM instructions. Besides, both SmartDagger~\cite{liao2022SmartDagger}, Neural-FEBI~\cite{he2023neural} and DeepInfer~\cite{zhao2023DeepInfer} are based on deep-learning models (e.g., deep neural networks) for information recovery. Therefore, their capability are quite limited to the model-scale and training datasets. For example, they suffers performance degradation when confronted with newly-emerging or rare contracts that are not exist in the training datasets. 
%


\para{Our work} In this paper, we propose \system{}, a novel framework designed to optimize the output of existing smart contract decompilers. In particular, \system{} aims at recovering \textit{function boundaries}, \textit{variable types}, and \textit{contract attributes} through a novel combination of static analysis (SA) and large language models (LLMs). The output of \system{} can facilitate a number of downstream tasks in program comprehension and security analysis. For example, eliminating false positives and false negatives for vulnerability detection. 

Our key insight is that 
(1) Software exhibits the natural patterns - programmers tend to utilize similar code structures, contract attributes, variable types, and function boundaries in comparable contexts~\cite{dramko2024taxonomy}. This repetitiveness enables predicting highly probable contract attributes, function boundaries and variable types for similar contexts.
(2) SA and LLMs can collaboratively enhance the output of existing decompilers. 
Specifically, the advantage of SA lies in its soundness, as it is accurate in handling the optimization targets with complex static constraints~\cite{he2018debin}. 
Consider the optimization of variable types as an example. The types of state variables (e.g., \textit{address}) are inherently related to the program expressions that access or modify them (e.g., \textit{msg.sender}). 
In the meantime, the strength of LLM lies in its completeness, as it possesses the flexibility to predict the optimization targets that lack static constraints~\cite{chen2022augmenting}.
For instance, with the remarkable capability of few-shot in-context learning, LLMs are flexible to predicting the types 
for rare local variables which are incapable of inferring their types based on the types of other variables.



\system{} extracts the program dependencies within the program through SA, and then leverages the superior generalization capabilities of LLMs to optimize the decompiled code after learning these program dependencies.
Firstly, \system{} extracts three types of dependencies from the decompiled code, including state dependency (e.g., read and write on state variable), control-flow dependency, and type dependency, to construct a fine-grained Dependency Graph (DG) (see Section~\ref{sec:DependencyExtraction}). The DG contains static domain knowledge (i.e., dependencies) necessary for decompiler output optimization, which facilitates the LLMs in capturing them across the entire smart contract.
Secondly, \system{} uses the DG to create prompts that include optimization target contexts (i.e., functions or variables), optimization result candidates (e.g., type or attribute), and chain-of-thoughts that represent inference steps of static analysis for optimization targets. 
By learning the prompts, LLMs enable the joint optimization on decompiler output with the advantages of both SA and LLMs (see Section~\ref{sec:OutputGeneration}).


A common challenge in LLM-adaptation is to handling its potential hallucination. In our task, LLMs unexpectedly alter the program behaviors of the original decompiled code,
and introduce inference errors that contradict common static domain rules (e.g., type knowledge)(see Section~\ref{sec:CorrectnessVerification}). 
As countermeasure,
\system{} conducts the rigorous correctness verification for the LLM output.
Specifically, \system{} utilizes symbolic execution and formal verification to validate the program-behavior equivalence between the original decompiled code and optimized code. In addition, \system{} integrates a set of static rules (e.g., type rules) to identify and reject inference errors.



\para{Evaluation} 
To evaluate the effectiveness of \system{}, we randomly select 500 functions from the largest open-source contract dataset~\cite{xblockcontract}, 
and manually labelled a dataset containing 456 pairs of source code and decompiler outputs of smart contract functions (44 functions encountered decompilation error). Noting that the dataset size is similar to those used in SOTA studies~\cite{hu2024degpt,ma2024combining}. The evaluation results indicate that, compared to the original decompiler output (i.e., Gigahorse~\cite{grech2019Gigahorse}), \system{} (with GPT-3.5) improves the precision and recall of function boundary identification by 20.30\% and 30.03\%. 
Further, \system{} significantly outperforming tool SOTA VarLifter~\cite{li2024varlifter}, effectively enhances the precision and recall of variable type inference by 13.51\% and 77.08\%.
Additionally, \system{} significantly outperforming SOTA tool SmartDagger~\cite{liao2022SmartDagger}, successfully improves the precision and recall of contract attributes by 44.69\% and 80.86\%
With the help of \system{}, 60.22\% of optimized codes can directly be recompiled using the Solidity compiler, whereas the original decompiled output can not support recompilation. 
We also prove that the optimized output of \system{} enhances the effectiveness of downstream tasks, i.e., vulnerability detection. For example, \system{} enhances the precision by 21.96\% and the recall by 38.00\% on detecting integer overflow, improve the precision by 16.67\% for contract attack identification.



In summary, this paper makes the following contributions:

\begin{itemize}

    \item We highlight the key limitations of current smart contract decompiler outputs that resistant various program analysis tasks in this domain.

    \item We propose \system{}, a novel framework for optimizing smart contract decompiler output in a generic manner. We propose a set of novel mechanisms
     that combine static analysis and large language models, to establish accurate and flexiable optimization in decompiler output.
    
    \item We perform extensive evaluation to show the effectiveness of \system{}. We demonstrate the efficacy of \system{} via three downstream tasks for vulnerability detection.

    \item To benefit future research, we release the artifact of \system{}, as well as the datasets~\footnote{{\publicUrl}}. 

\end{itemize}
    

\section{Background and Motivation}
\label{sec:background}

\subsection{Smart Contract Decompiler}
\label{sec:decompilation}


Smart contract is a specific type of program running on the blockchain~\cite{su2023hybrid}, which  has enabled a wide range of application scenarios in the digital world~\cite{yao2024distributed}, such as Decentralized Finance~\cite{liao2024smartaxe}, Supply Chain Management~\cite{wang2024blockchain}, and Internet of Things~\cite{al2024blockchain}. Recently, the investigation report show that more than 99\% smart contracts do not disclose their source code, smart contract decompilation become increasingly important. 

Decompilation involves the recovery of variables, functions, and control-flow abstractions through various program analysis methods. Further, it then utilizes heuristic rules to synthesize these abstractions, thereby reconstructing the high-level code representation, commonly referred as decompiled code~\cite{chen2022augmenting}. Recently, an investigation report reveals that over 99\% of smart contracts do not make their source code publicly available, with only their bytecode being accessible~\cite{ContractStatistic}. Therefore, smart contract decompilation is becoming increasingly important.

Smart contract decompilation faces significant challenge in accurately restoring bytecode to its original source code~\cite{jaffe2018meaningful}. 
This challenge is primarily caused by two factors: (1) the compiler may lose crucial information about the original program. For example, the lexical parsing stage of the compiler fails to propagate \textit{variable types}, \textit{variable names}, \textit{annotation} and \textit{function boundary} into the bytecode~\cite{grech2018Madmax}; and (2) the heuristic rules employed by existing decompilers exhibit low coverage rates~\cite{hu2024degpt}, This which make the decompiler difficult to adapt to varying contract programming styles or patterns.




\begin{figure*}[t]
\includegraphics[width=7.2in]{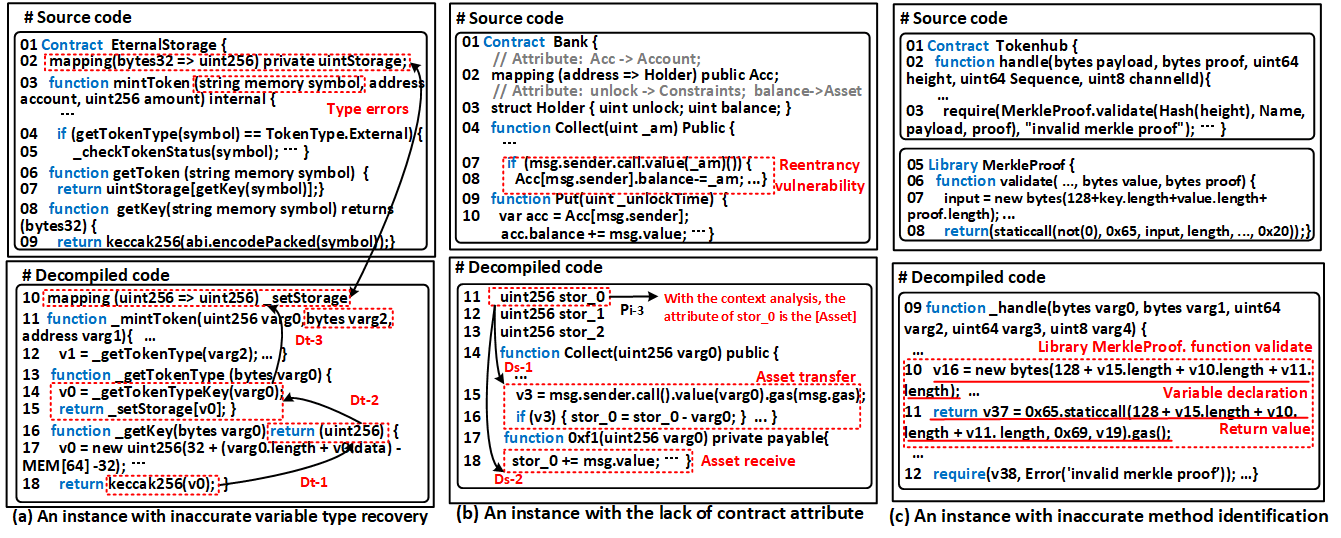}
\caption{Three motivating examples for illustrating the limitations of current decompiler output. }
\label{motivatingexample}
\end{figure*}

\subsection{Deficiencies in Decompiler Output}
\label{sec:deficiency}
The current decompiler output exhibits certain deficiencies that obstruct effective comprehension and analysis of the target program.
Below we take smart contracts of Fig.~\ref{motivatingexample}, which contain source code and corresponding decompilation code, as the instances for illustrating the deficiencies.


\para{Deficiency-1. Inaccurate Variable Type Recovery} 
Limited by the insufficiently generalized heuristic rule, the SOTA decompilers are inaccurate in recovering the variable type. This deficiency is evident in the example shown in Fig.~\ref{motivatingexample}(a). 
In the source code, the state variable \textit{uintStorage} is a mapping type that maps \textit{bytes32} to \textit{uint256} (line 2). However, the decompiler incorrectly assigns it as a mapping \textit{uint256} to \textit{uint256} (line 10). Additionally, while decompilers narrow down the range of possible types for the parameter \textit{varg2} and assign it with \textit{bytes} (line 11), the correct type of this parameter is actually a \textit{string array} (line 3). Obviously, the SOTA decompilers commit errors in type inference. 
 The notable efforts in this domain are SigRec~\cite{chen2022SigRec}, VarLifter~\cite{li2024varlifter} and DeepInfer~\cite{zhao2023DeepInfer}. SigRec and DeepInfer focus primarily on inferring function signatures and recovering parameter types within those signatures. Since these tools concentrate on a subset of variable types, they are insufficient to address the broader deficiency of type recovery (e.g., variable \textit{uintStorage}). VarLifter exhibit low coverage rates, because it heavily relied on
pre-defined heuristics according to EVM instructions. In addition, the inaccuracies in variable type recovery result in significant difficulty in identifying specific vulnerabilities, such as overflow vulnerabilities~\cite{lai2020static}.

\para{Deficiency-2. Lack of Contract Attribute} 
Due to the limited storage of blockchain, smart contract utilizes state variables (i.e., global variables) to record the key contract attributes (e.g., asset, identity, router).
While contract attributes are crucial for the vulnerability analysis task, they are exceptionally difficult to be inferred from decompiled code~\cite{liao2022SmartDagger}. We take an example in Fig. 1(b) for illustration. Contract \textit{Bank} exhibits a Reentrancy vulnerability because it records the asset balance after transferring assets.
An adversary could exploit this by re-entering the contract to transfer assets without recording the change of asset balance (line 7-8). Thus, correctly identifying the state variable \textit{Acc.balance} as an \textit{Asset} attribute is critical for detecting this vulnerability.
While the contract attribute is explicitly stated in the source code through meaningful variable names (i.e., \textit{Acc.balance}), it is totally missed at the bytecode
level and is labeled as \textit{store\_0} in its decompiled code (line 11). Hence, inferring the contract attribute for the variable \textit{store\_0} in the decompiled code becomes a formidable challenge. Without the contract attribute, it is difficult to pinpoint the vulnerability.
Prior work (i.e., SmartDagger~\cite{liao2022SmartDagger}) trains a neural machine learning model using a corpus of 1,200 smart contracts to recover contract attributes for state variables within smart contracts. However, due to limitations in model capacity and the size of the training set, SmartDagger degrade its accuracy of contract attribute prediction when encountering with newly-emerged or rare contracts.


\para{Deficiency-3. Inaccurate function Identification}
SOTA decompilers are notably ineffective in identifying functions within smart contracts due to their reliance on low-coverage heuristic rules. For instance, the Gigahorse~\cite{grech2019Gigahorse} decompiler identifies functions by detecting call sites that are invoked recurrently. The heuristic rules result in poor performance when encountering complex functions, particularly inherited functions. In contrast to languages like C++ and Java, where inheritance is clearly delineated, smart contract inheritance involves embedding all base subcontracts ($B_1, B_2,...,B_n$) as code blocks directly into the inheriting contract $A$, without retaining explicit call information~\cite{Solidity}. Hence, the heuristic rules of Gigahorse failed in this case. As can be seen in Fig.~\ref{motivatingexample}(c), the inherited function \textit{validate} is explicitly stated in the source code (line 6-8), but it is absent at the bytecode level. Our manual investigation reveals that the \textit{validate} function is integrated into the \textit{handle} function (lines 10-11). Consequently, SOTA decompilers like Gigahorse are unable to reconstruct such inherited functions, which are critical for downstream program analysis tasks such as cross-contract call flow analysis~\cite{yang2024uncover,liao2022SmartDagger} and
component analysis~\cite{sun2023Demystifying}.





\begin{figure*}[t]
    \centering
    \begin{minipage}[t]{.5\textwidth}
        \centering
        \includegraphics[width=0.95\linewidth]{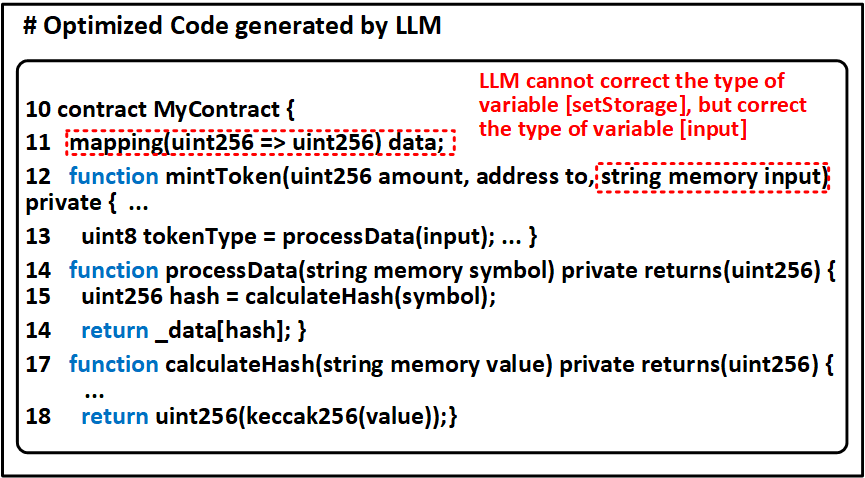}
        \caption{The decompiled code optimized by LLM for the \\ instance in Fig.~\ref{motivatingexample}(b).}
        \label{static analysis}
    \end{minipage}%
    \begin{minipage}[t]{.5\textwidth}
        \centering
        \includegraphics[width=0.95\linewidth]{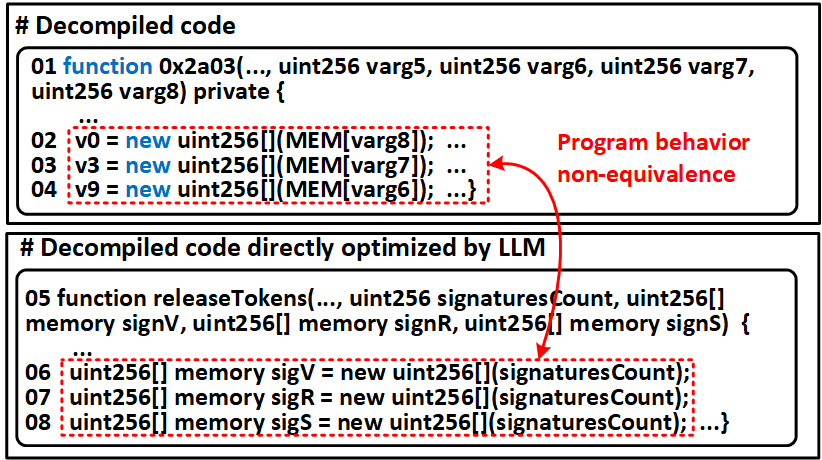}
        \caption{An optimization error reported by LLM inference in terms of program-behavior non-equivalence.}
        \label{optimization error}
    \end{minipage}
\end{figure*}

\subsection{Our solution}
\label{sec:solution}

We propose our strategies for optimizing the decompiler output, combining with a discussion on potential solutions for the deficiencies presented in motivating examples of Fig.~\ref{motivatingexample}.

\para{Strategy-1: Using SA to extract control-flow dependency, type dependency and state dependency}
First, type dependency is crucial for the variable type recovery. For example, the type error for variable \textit{uintStorage} can be corrected by using SA to infer type dependencies, as illustrated in Fig.~\ref{motivatingexample}(a). Initially, the output of the predefined function \textit{keccak256} is of type \textit{bytes32} ($D_{t_1}$). Subsequently, the type of variable $v0$ depends on the return value type of the function \textit{\_getTokenKey}, which is also \textit{bytes32} ($D_{t_2}$).
 Finally, the key type of the variable \textit{setStorage} depends on the type of $v0$ ($D_{t_3})$. Consequently, the type of \textit{setStorage} should be corrected to \textit{mapping(bytes32=\>uint256)}.
Second, control-flow dependency is critical for the function boundary identification. For instance, 
as shown in Fig.~\ref{motivatingexample}(c), while static constraints (i.e., call information) are unavailable in function \textit{validate}, we can manually infer its boundaries by analyzing the contextual control flow, i.e., the function typically declares a new variable at the beginning (line 10) and returns the value at the end (line 11).
Third, state dependency is important for contract attribute inference.
We take the motivating example in Fig.~\ref{motivatingexample}(b) as an instance for illustration.
By analyzing the dependency of the state variable, we can find that state variable \textit{stor\_0} is written in line 16 ($D_{s_1}$) and line 18 ($D_{s_2}$). 
By manually analyzing their context usage, we can infer that they are utilized for asset transfer and asset receive, respectively. Hence, we can infer the attribute for \textit{stor\_0} as the \textit{[Asset]}.

\para{Strategy-2: Empowering LLMs to enrich semantics}
With the remarkable capability of few-shot in-context learning~\cite{li2025channel,wang2025preformer}, LLMs are flexible to predicting the optimization target (i.e., variables or functions), which are rarely encountered in the program and often incapable of inferring them based on other variables or functions.  We take Fig.~\ref{motivatingexample}(a) again as an instance for illustration.
The type error for variable \textit{input} can be corrected by using LLMs for prediction.  Fig.~\ref{static analysis} presents code optimized by LLM.  Benefiting from the extensive training set and model capabilities, LLM can assign the variable \textit{input} with \textit{string memory} correctly (line 12). 
In addition, owing to its remarkable model capacity, LLMs are more effective and generalizable in learning static domain knowledge (i.e., dependencies) for optimization predictions, compared to SA's manually crafting heuristic rules based on the same knowledge~\cite{peng2023generative}. For instances, instead of manually summarizing heuristic rules, we can utilize LLMs to learning the control-flow context for function boundary prediction, and to analyze the contextual usage of state variables to infer their contract attributes.

Inspired by the above analysis, our objective is to harness the advantages of both SA and LLMs to optimize decompiler output. 
Therefore, SmartHalo extracts three types of program dependencies including type dependency, state dependency and control flow dependency 
within the program through SA. Subsequently, \system{} leverages the superior generalization capabilities of LLMs to optimize the decompiled code after learning these program dependencies.

\begin{figure*}[t]
\centering
\includegraphics[width=7.2in]{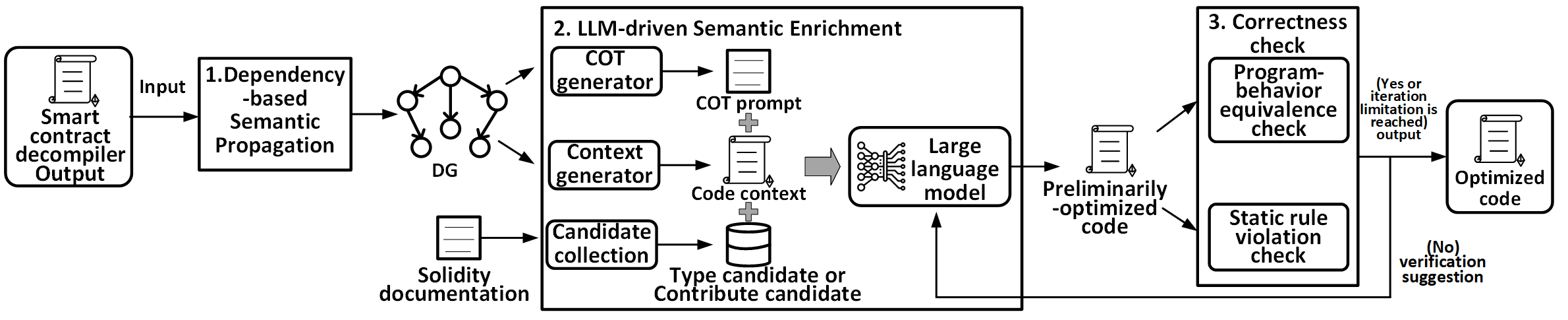}
\caption{The overview of \system{}.} 
\label{fig:overview}
\end{figure*}

\section{Design of \system{}}
\label{sec:overview}




To overcome the deficiencies, we propose a novel decompiler output optimization framework, called \system{}, based on hybrid SA and LLM. 
\system{} takes the decompiler output as input, and finally outputs the optimized code. Fig.~\ref{fig:overview} shows the overview of \system{}, which integrates the following mechanisms to ensure its effectiveness.

\para{Dependency-based  Semantic Extraction}
In this mechanism, \system{} aims to extract critical semantics from the decompiled code through a fine-grain dependency analysis, which is then propagated into the LLM for further processing.
To this end, our work complements and extends previous research~\cite{zhao2023DeepInfer,chen2022SigRec,li2024varlifter,liao2022SmartDagger} by considering two new types of dependencies: namely, state dependency and type dependency. Specifically, \system{} analyzes three types of dependencies as the crucial program semantics, subsequently utilizing them to construct a Dependency Graph (DG).
The extracted semantics (i.e., dependencies) represent the critical static domain knowledge necessary for decompiler output optimization,
which facilitates the LLMs in capturing them across the entire
smart contract (See Section~\ref{sec:DependencyExtraction}).

\para{LLM-driven Semantic Enrichment} With the semantic re-presentation (i.e., DG) propagated by the above mechanism, \system{} constructs inference prompts to guide LLM to enrich the semantic of decompiler output. 
Specifically, according to the DG, \system{} utilizes our designed generator to  
constructs the context for optimization targets (i.e., functions or variables), for determining the code range related to optimization targets. Subsequently, \system{} further generates chain-of-thoughts that represent inference steps of static analysis for optimization targets, based on the dependencies among the DG, for teaching LLMs how to infer variable types, contract attributes and function boundaries. In addition, \system{} collects the candidates for variable types and contract attributes from the Solidity documentation~\cite{Solidity}. Lastly, LLM learns from
the above three types of prompt, and optimizes  variable types, contract attributes and function boundaries for decompiled code (See Section~\ref{sec:OutputGeneration}).









\para{Correctness Verification}
However, verifying correctness for the optimized code is far from straightforward.
First, due to the hallucination that commonly exist in LLMs, LLMs generates the random and unpredictable outputs that unexpectedly alter the program behaviors of the original decompiled code.
For example, Fig.~\ref{optimization error} illustrates a program behavior non-equivalence error caused by LLM optimization.
In the original decompiled code, variables \textit{v0}, \textit{v3} and \textit{v9} depends on parameters \textit{varg6}, \textit{varg7} and \textit{varg8}, respectively (line 02-04). However, in the optimized code, variables \textit{sigV}, \textit{sigR} and \textit{sigS}  incorrectly depend on the same parameter \textit{signaturesCount} (line 06-08).
Second, LLMs also introduce errors that contradict common static domain rules (e.g., type rules), further complicating correctness verification. 
We take Fig.~\ref{static analysis} again as an instance for illustrating such type of errors. In line 18, the value returned by the predefined function \textit{keccak256} is type of \textit{bytes32}. According to Solidity documentation~\cite{Solidity}, \textit{bytes32} cannot be directly converted into \textit{uint256}, result in convertion errors.



\system{} implements two specific checks for establishing correctness verification. In the program-behavior equivalence check, \system{} utilizes symbolic execution and formal verification to verify the program-behavior equivalence between the original decompiled code and optimized code. In the static rule violation check, \system{} summarizes a set of static rules that are related to types. 
Note that static rule violation check donot consider function boundaries and contract attributes because there are no explicit static rules related to them in smart contract~\cite{Solidity}.
By applying these static rules, \system{} rejects the inference errors. (see Section~\ref{sec:CorrectnessVerification})









\section{Approach Details}
\label{sec:methodology}

\subsection{Dependency-based Semantic Extraction}
\label{sec:DependencyExtraction}

\para{Basic control-flow analysis}
To enable control-flow analysis at the level of decompiled code, we develop a Solidity-oriented control-flow analysis component for SmartHalo, which provides both code parsing and control/data-flow graph construction. Built on a Tree-sitter architecture, the component performs analysis without requiring compilation, which eliminates environment configuration and build overhead. Such property makes the  component particularly suitable for processing decompiled pseudocode. Specifically, the control-flow analysis component (i) employs Tree-sitter~\cite{Tree-sitter} to construct a concrete syntax tree for each decompiled code file, (ii) transforms the syntax tree into a three-address intermediate representation (IR), and (iii) leverages this IR together with function-call information to produce the corresponding control-flow and data-flow graphs.

\para{Contract Dependency Identification} \system{} focus on identifying dependencies including the type dependency, state dependency, and control-flow dependency.

\begin{itemize}
    \item \textit{Type dependency.} Type dependency refers to the dependency that the type of variable $v_a$ correlates to another variable $v_b$ or a specific expression $e_b$. Fig.~\ref{fig:sytax} shows the syntax of all the expressions that generate types in smart contracts. Given the decompiled code, \system{} identifies the type dependency according to these syntaxes effectively.


    \item \textit{State dependency.} State dependency refers to the dependency between different state variables or dependency between state variables and expressions, such as read and write dependency on state variables. \system{} utilize SmartState~\cite{liao2023SmartState}, a SOTA state dependency analyzer to identify state dependency from given smart contracts. 
\end{itemize}


\para{Dependency Graph Construction}
Subsequently, \system{} proposes the dependency graph (DG) as the representation of critical semantics for smart contracts. 

The DG constructed by \system{} can be represented as a tripe $G_c=(N_c,E_c,X_e)$. Specifically, \system{} encodes the following information: (1) the nodes of DG can be denoted as a set of all variables and related expressions in the decompiled code. Here, we use $N_v$ to represent the variable nodes, and leverage $N_e$ to represent the expression nodes. 
Hence, we have $N_c:=\left\{ N_v \cup N_e\right\}$; (2) the edges of DG can be represented as a set of control-flow dependency edges $E_d$, state dependency edges $E_s$ and type dependency edges $E_t$. Similarly, we have $E_c:=\left\{ E_d \cup E_s \cup E_t\right\}$. (3) $X(E_{c}) \rightarrow \left\{DFD, SD, TD \right\}$ is a labeling function that maps an edge to one of the above three dependencies.

\begin{figure}[h]
\footnotesize
\begin{align*}
e \in Expr  ::= & v~ |~ c~ |~ e~ blop~ e~ |~ e~ numop~ e~ |~ e~ cmpop~ e~ |  \\
~~~~~~~~~~~~~~  & e~ bitop~ e~ |~ (e, \dots, e) \ |~ [e, \dots, e]~ |~ e(e, \dots, e)~ | \\
~~~~~~~~~~~~~~  & e[e]~ |~ e[e:~ e] ~ |~ e.v  \\
\end{align*}
\vspace{-0.3in}
\caption{The syntax of expression for typing in Solidity.} 
\label{fig:sytax}
\end{figure}


After generating control-flow graph, \system{} constructs the DG by incrementally adding
state dependency edges and type dependency edges to the control-flow graph. Specifically, firstly, \system{} searches the whole control-flow graph to locate the variables and expressions that are related to type dependency. Further, \system{} utilizes SmartState to identify the variables and expressions that are related to state dependency. Finally, \system{} finds out the sources and targets among all these variables and expressions, and uses the direct edge to connect them in pair. 


\subsection{LLM-driven Semantic Enrichment}
\label{sec:OutputGeneration}


In this subsection, \system{} first generates domain-aware inference prompts according to the DG. Then, \system{} utilizes the inference prompts to guide the LLM to optimize the decompiler output. \system{} considers three types of inference prompts, including code contexts, inference candidates, and Chain-of-Thought (COT) prompts.




\begin{table*}[b]
\footnotesize
\centering
\caption{Chain-of-Thought Prompt Template for LLM-driven semantic enrichment}. 
\setlength{\tabcolsep}{0.4mm}{
\begin{tabular}{c|l|l}
\hline
\multicolumn{1}{l}{Category}                                                          & \multicolumn{1}{c}{Type}                                                       & \multicolumn{1}{c}{Template}                                                                                                                         \\ \hline
\multirow{5}{*}{\begin{tabular}[c]{@{}c@{}}Type \\  dependency\end{tabular}}        & Variable $\to$ Variable                                                   & \begin{tabular}[c]{@{}l@{}}The type of variable {[}NAME{]} is consistent with the type of variable {[}NAME{]}\end{tabular}                          \\
                                                                                      & Type $\to$ Expression                                                     & \begin{tabular}[c]{@{}l@{}}The value of predefined function/operands{[}NAME{]}  of expression {[}STATEMENT{]} is type of {[}TYPE{]}\end{tabular}   \\
                                                                                      & Type $\to$ Variable                                                       & The variable of {[}NAME{]} is assigned from {[}TYPE{]}                                                                                               \\
                                                                                      & Expression $\to$ Variable                                                 & \begin{tabular}[c]{@{}l@{}}The type of variable {[}NAME{]} depends on  expression {[}STATEMENT{]}\end{tabular}                                     \\
                                                                                      & Variable $\to$ Expression                                                 &    \begin{tabular}[c]{@{}l@{}}The operand(s)/target(s)/key(s)/value(s)
of expression [STATEMENT] is/are [TYPE]. \end{tabular}                                                                                                                                                  \\ \hline
\multirow{3}{*}{\begin{tabular}[c]{@{}c@{}}State \\ dependency\end{tabular}}       & State variable $\to$ State variable                                       & \begin{tabular}[c]{@{}l@{}}The attribute of state variable {[}NAME{]} is  correlated to the attribute of state variable {[}NAME{]}\end{tabular} \\
                                                                                      & State variable $\to$ Expression                                           & \begin{tabular}[c]{@{}l@{}} The attribute of state variable {[}NAME{]} is correlated to the context usages of Expression  {[}STATEMENT{]} (Write)\end{tabular}                                         \\
                                                                                      & Expression $\to$ State variable                                           & \begin{tabular}[c]{@{}l@{}}The attribute of state variable {[}NAME{]} is correlated to the context usages of Expression  {[}STATEMENT{]} (Read)\end{tabular}                                            \\ \hline
\multirow{4}{*}{\begin{tabular}[c]{@{}c@{}}Control flow\\ dependency\end{tabular}} & \begin{tabular}[c]{@{}l@{}}
 Call Site \end{tabular}                                                                        & Expression [STATEMENT] seems to be a call site to another function. \\

& Modifier                                                                         & Expressions with a range from [STATEMENT] to [STATEMENT] seems to be a modifier function. \\

&Return Value  & Expression [STATEMENT] is used to return the value. Please determine whether it is the end point.                                                                                                                                                     \\ & Variable Declaration              & Expression [STATEMENT] is used to declare the variable. Please determine whether it is the start point.             
                                                                                                                                                                                          \\ \hline
\end{tabular} }

{Note: [NAME] indicates the names of variable nodes, [STATEMENT] indicates the statements of expression nodes, [TYPE] indicates the types of variable nodes. [USAGES] indicates the usages for the variable nodes.}

\label{table.COT}
\end{table*}

\para{Code Context} 
\system{} adapts different strategies to generate code snippets for variables and functions. The strategy for variables applies to optimizing variable type and contract attribute, and another strategy for functions is used to optimize function boundaries.

$\bullet$~\textit{Variables}. To pinpoint the context related to target variables precisely, \system{} utilizes code slicing to identify code context fragments from the decompiled code for target variables, based on the DG. 
To this end, \system{} first generates the slicing DG based on the original DG. On the original DG, \system{} starts from target variable nodes and performs the forward and backward traversals (i.e., the same or opposite direction to DG edges). In this way, \system{} finds out all the nodes that have the dependency on target variables and generates a slicing subgraph of DG.
With the slicing DG, \system{} combines the corresponding expressions together to generate a code snippet as the context prompt.

$\bullet$~\textit{functions.} To generate the context prompt for target functions, \system{} searches the DG to find out the call chains that target functions lie in. Then, \system{} combines the corresponding functions in these call chains together to generate a code snippet as the context prompt.



\begin{figure*}[t]
\centering
\includegraphics[width=7.4in]{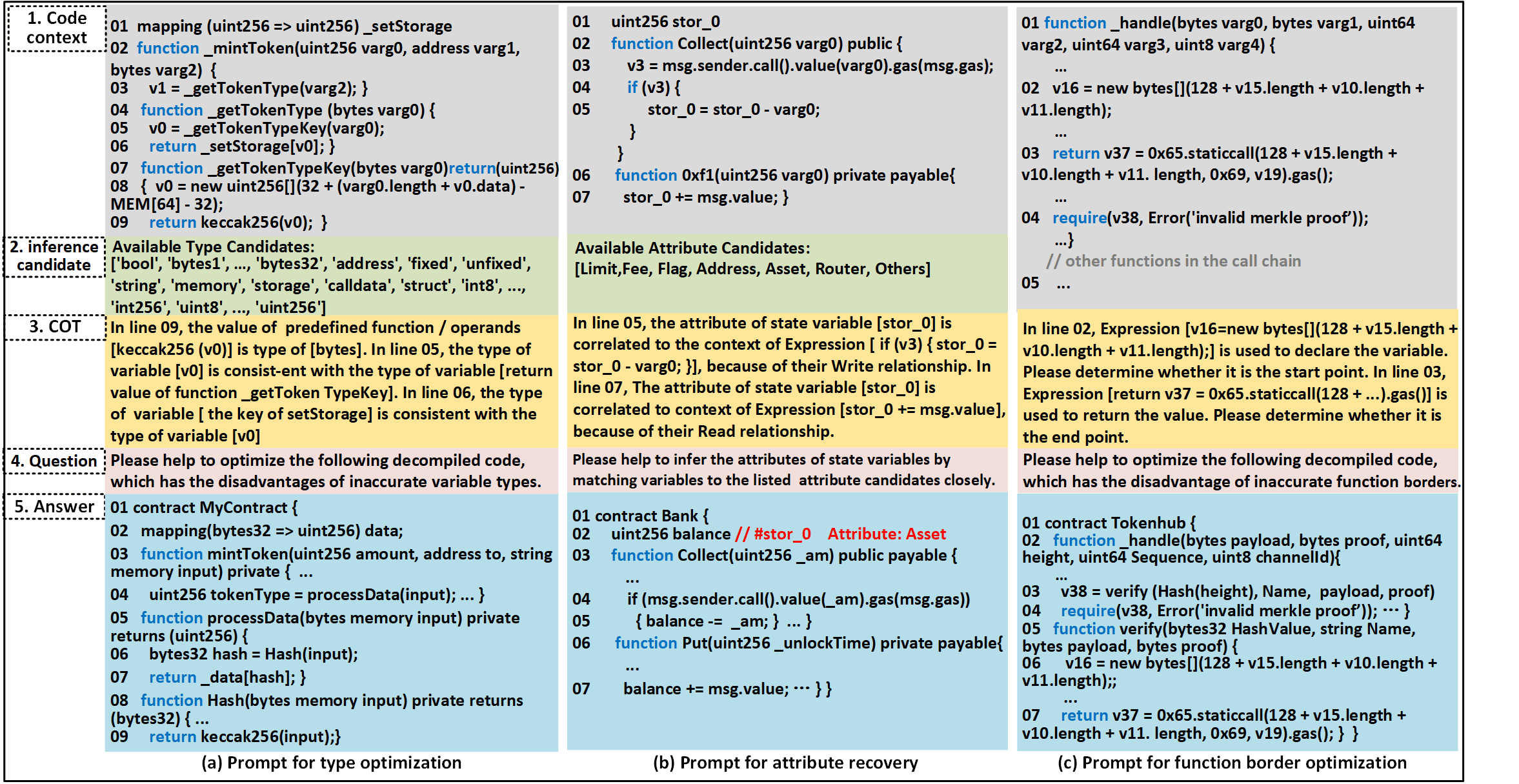}


\caption{The prompt for the instance in Figure~\ref{motivatingexample}.}
\label{LLM inference}
\end{figure*}

In this way, \system{} narrows down the original decompiled code to the much smaller code snippet, which contains the information only related to decompiler output optimization of target variables or functions. 
We present the code context generation for the instance of Fig.~\ref{motivatingexample}, as highlighted in the gray part of Fig.~\ref{LLM inference}. For type optimization, \system{} identifies all the expressions that have type dependency related to variable \textit{setStorage}, and integrates them as the code context prompt.
For attribute recovery, \system{} extracts the corresponding expressions that have state dependency on state variable \textit{stor\_1}, and aggregates them as the code context prompt.
Finally, \system{} extract the functions from the call chain that function $\_handle$ lies in to form the code context prompt.

\para{Inference Candidate}
Subsequently, \system{} prompts the LLM with the candidates of variable type and contract attribute. In this way, \system{} can help the LLM narrow down the selection range to improve the accuracy.

\textit{Variable type candidates.} 
Solidity is a statically typed language that provides a specific number of built-in types. In addition, the user-defined value type in Solidity is actually an alias that creates a zero cost abstraction over an elementary value type, so \system{} directly utilizes the elementary value type to represent it. According to Solidity documentation~\cite{Solidity}, we collect the variable type candidates, as can be seen in the inference candidates (highlighted in green part) of Fig.~\ref{LLM inference} (a).

\textit{Contract attribute candidates.}
In contrast to variable type, contract attribute is far from straightforward and should be summarized by representative smart contract dataset. We summarize the contract attribute candidates as a set, 
as can be seen in the inference candidates (highlighted in green part) of Fig.~\ref{LLM inference} (b). And the detail of summarizing process are illustrated in Section~\ref{sec:Implementation}



\begin{figure*} [b]

\begin{flalign}
& \frac{}{\pi \vdash c:  \theta  } \tag{Constant} & \\
& \frac{\pi \vdash e_1:  \theta_1, \quad \pi \vdash e_2:  \theta_2,  \quad \widetilde{\theta }= \left \{ bool, int \right \}  }{\pi \vdash e_1  [bitop] e_2 : \theta \wedge \widetilde{\theta }, \quad \pi \vdash e_1: \theta_1 \wedge \widetilde{\theta }, \quad \pi \vdash e_2: \theta_2 \wedge \widetilde{\theta }} \tag{LShift, RShift} & \\
&\frac{\pi \vdash e_1:  \theta_1, \quad \widetilde{\theta }= \left \{ bool, int \right \}, \quad \pi \vdash e_2:  \theta_2,  \quad {\theta }'= getMorePreciseType(\theta_1 \wedge \widetilde{\theta }, \quad \theta_2 \wedge \widetilde{\theta })  }{\pi \vdash e_1 [numop] e_2 :  {\theta}',  \quad \pi \vdash \theta_1 \wedge \widetilde{\theta }, \pi \vdash \theta_2 \wedge \widetilde{\theta }} \tag{Numeric Operations}& \\
&\frac{\pi \vdash e_1:  \theta_1, \quad \pi \vdash e_2:  \theta_2,  \quad {\theta }'= \left \{ \Gamma, Array, Tuple \right \}   }{\pi \vdash e_1 [cmpop] e_2 :  bool,  \quad \pi \vdash e_1: \theta_1 \wedge \theta_2 \wedge \widetilde{\theta }, \quad \pi \vdash e_2: \theta_1 \wedge \theta_2 \wedge \widetilde{\theta }} \tag{Lt, LtE, Gt, GtE}& \\
&\frac{\pi \vdash e_1:  \theta_1, \quad ..., \quad \pi \vdash e_n:  \theta_n   }{\pi \vdash (e_1,...,e_n):Tuple[\theta_1,...,\theta_n], \quad \pi \vdash (e_1,...,e_n):Array[\theta_1,...,\theta_n] } \tag{Tuple, Array}& \\
&\frac{\pi \vdash e:  \theta, \quad \widetilde{\theta }= \left \{ str, bytes \right \},  \quad {\theta }'= getElementType(\theta_1 \wedge \widetilde{\theta })  }{\pi \vdash [for] v [in] e :  {\theta}',  \quad \pi \vdash e:\theta \wedge \widetilde{\theta }} \tag{Comprehension}& \\
&\frac{\pi \vdash e_1:  \theta_1, \quad \pi \vdash e_2:  \theta_2  }{\pi \vdash e_1  [blop] e_2 : {\theta }', \quad Union[\theta_1,\theta_2]}  \tag{Boolean Operation}& \\
&\frac{\pi \vdash e_1:  \theta_1, \quad \pi \vdash e_2:  \theta_2,  \quad \widetilde{\theta }= \left \{ bool, int, byte \right \}  }{\pi \vdash e_1  [bitop] e_2 : \theta \wedge \widetilde{\theta }, \quad \pi \vdash e_1: \theta_1 \wedge \widetilde{\theta }, \quad \pi \vdash e_2: \theta_2 \wedge \widetilde{\theta }} \tag{Bitor, BitAnd, BitXor}&  \\
&\frac{\pi \vdash e_1:  \theta_1, \quad \pi \vdash e_2:  \theta_2  }{\pi \vdash e_1  [cmpop] e_2 : bool}  \tag{Eq,NotEq,Is,IsNot}& \\
& \frac{\pi \vdash e_1:  \theta_1, \quad \pi \vdash e_2:  \theta_2,...\quad \pi \vdash e_n:  \theta_n,  \quad \widetilde{\theta }= \left \{ Callable[[\theta_1,\theta_2,...,\theta_n],\theta] \right \}, \quad {\theta}'=getReturnType(\theta \wedge \widetilde{\theta })  }{\pi \vdash e(e_1,...,e_n):\theta  } \tag{Call}& \\
&\frac{\pi \vdash e_1:  \theta_1, \quad \pi \vdash e_2:  \theta_2, \quad \widetilde{\theta_1 }= \left \{ str, bytes \right \}, \quad \widetilde{\theta_2 }= \left \{ int, bool \right \}, \quad {\theta }'= getElementType(\theta_1 \wedge \theta_2)  }{\pi \vdash e_1[e_2]:{\theta}', \quad \pi \vdash e_1:\theta_1 \wedge \widetilde{\theta_1} , \quad \pi \vdash e_2:\theta_2 \wedge \widetilde{\theta_2}} \tag{Slice}&
\end{flalign}
\caption{The static rules for type violation check.}
\label{fig:StaticRule}

\end{figure*}

 \para{Chain-of-Thought Prompt} 
The COT prompts are a series of intermediate reasoning steps in static analysis.
\system{} prompts the LLM with COT Prompt, for teaching LLM how to reason the correct function boundaries, variable type and contract attribute from the perspective of static domain knowledge.
 

\system{} utilizes the slicing DG of the code slicing phase to generate the COT prompts. Given the slicing DG, \system{} organizes the nodes according to the distance between the nodes and the target variable node (i.e., hop counts). Specifically, the hop count of the target variable node is set as 0, and the hop counts of other nodes are calculated by their distance. \system{} translates each hop in the DG into a sentence of dependency description.
Inspired by prior work~\cite{peng2023generative}, we summarize the effective COT prompt templates, as shown in Table~\ref{table.COT}. 
As can be seen, each optimization tasks involve at least one kind of COT prompt template.
Specifically, variable type optimization relies on the categories of both type dependency and control-flow dependency. 
Further, variable type optimization depends on the categories of both static dependency and control-flow dependency.
And function boundary optimization only requires the category of control-flow dependency.
To generate the COT prompt, \system{} fills the corresponding variable [TYPE], variable [NAME], [USAGE] and [STATEMENT] into the templates.

To illustrate the COT prompt generation, we present the COT prompt for the instance of Fig.~\ref{motivatingexample},  as highlighted in yellow part of Fig.~\ref{LLM inference}. For type optimization, as discussed above, there are three type dependency edges (i.e., $D_{t_1}$, $D_{t_2}$, $D_{t_3}$) for the instance of Fig.~\ref{motivatingexample}(a). As shown in Fig.~\ref{LLM inference}(a), \system{} utilizes $Expression \rightarrow Variable$ template to describe $D_{t_1}$, and leverages $Variable \rightarrow Variable$ template to describe $D_{t_2}$ and $D_{t_3}$. 
The sentences of all the corresponding edges are connected together according to the order of type dependency, to form the COT prompt for type optimization. 

For contract attribute recovery, as mentioned above, there are two dependency edges (i.e., $D_{s_1}$, $D_{s_2}$) for the instance of Fig.~\ref{motivatingexample}(b). As shown in Fig.~\ref{LLM inference}(b), \system{} utilizes $State variable \rightarrow Expression$ template to describe $D_{s_1}$ and and $D_{s_2}$. The sentences of these edges are aggregated together as the COT prompt for attribute recovery.

Fig.~\ref{LLM inference}(c) shows the COT prompt for function boundary optimization on the instance of Fig.~\ref{motivatingexample}(c). \system{} utilizes $Return-Value$ and $Variable-Declaration$ templates to describe statements that are likely to be the start and end points of a function.

\subsection{Correctness Verification}
\label{sec:CorrectnessVerification}


\para{Program-behavior Equivalence Check}
\system{} aims to check the program-behavior equivalence between functions of original decompiled code and functions of optimized decompiled code. We use $m$ and $m'$ to represent the two versions of the same function in the original decompiled code and optimized decompiled code.


\system{} conducts the symbolic execution on functions $m$ and $m'$ for generating symbolic summary $s$ and $s'$. If functions $m$ and $m'$ are equivalent in terms of program behaviors, their symbolic summary $s$ and $s'$ are also equivalent in terms of functionality. An equivalence assertion is a first-order logic formula $\Phi$ that is used to assert such equivalence~\cite{person2008differential}.
\begin{equation}
    \Phi =\neg (s\Leftrightarrow s^{'})
\end{equation}




\system{} utilize formal verification method to analyze symbolic summaries $s$ and $s'$, and determine whether equivalence assertion $\Phi$ is satisfied. If $\Phi$ is satisfied, function $m$ is non-equivalent to function $m'$ in terms of program behaviors, otherwise, they are equivalent to each other. Further, \system{} integrates the formal verification method that relies on SMT solver (i.e., Z3 solver~\cite{moura2008Z3}), to find out satisfying assignments for equivalence assertion $\Phi$. 

Note that since the symbolic summaries generally contain non-linear constraints, it is difficult for the Z3 solver to solve them. Such non-linear constraints commonly result in the solving failure. We eliminate most of the solving failure by utilizing methods proposed in ARDiff~\cite{badihi2020ARDiff}.

\para{Rule-based Type Violation Check}
\system{} traverse throughout the optimized code, then leverages violation rejection rules to identify and reject incorrect variable types predicted by the LLM. 

Here, we summarize the violation rejection rules integrated by \system{} in Figure~\ref{fig:StaticRule}. Each violation rejection rule is composed of two parts, including specific premises (contents above the line) and conclusions (contents below the line). They are organized as follows.
\begin{equation}
    \pi \vdash e : \theta
\end{equation}
where $\pi$ represents the context that contains lists that assign
variable types to expression patterns. In this form, $e$ refers to the expression,
and we use $e_1, ..., e_n$ to represent different expressions. $\theta$ is the variable types. We use $\theta_1, ..., \theta_n$ to represent different variable types. A rule in this form is called a judgment or assignment. Our goal is to get the context $\pi$ that assigns variable types to all the variables in code.





Finally, the incorrect optimization results predicted by LLM would be integrated with verification suggestions to form the violation information. \system{} retransmits the violation information into LLM to infer the correct function boundaries, variable types, and contract attributes anew. 
Further, the interaction can be iterated continually between LLM-driven semantic enrichment and correctness verification, until \system{} amends all the errors identified by correctness verification, or a maximum iteration limit is reached.




\section{Evaluation}
\label{sec:eval}

\para{Research Questions} 
We summarize the following research questions for evaluating \system{}.
\begin{enumerate} [RQ1.]
    \item How effective is \system{} in terms of decompiler output optimization?
    \item How does \system{} perform compared to other state-of-the-art mechanisms?
    \item How effective are individual components of \system{} in terms of helping \system{} improve the precision of decompiler output optimization?
    
    \item Can \system{} generalize to real-world complex contracts, different decompilers or different LLMs?
    \item What is the efficiency of \system{} in terms of both runtime and monetary costs?
    
    \item How effective is the output of \system{} in terms of helping downstream program analysis task?
\end{enumerate}




\subsection{Implementation and Evaluation Setup}
\label{sec:Implementation}

\para{Implementation}
\system{} was implemented with around 1,799 lines of code in Python 3.8.10.
We utilize ChatGPT (GPT-3.5) to support the LLM component of \system{}, due to the significantly expensive overhead of GPT fee. 
Note that \system{} is designed to be model-agnostic and unlimited to a certain large language model, it offers the flexibility to integrate alternative large language models.
 All evaluation experiments were conducted on an Ubuntu 20.04 server equipped with an Intel i9-10980XE CPU (3.0GHz), an RTX 3090 GPU, and 250GB of RAM.



\para{Evaluation Dataset and Ground-truth Establishment}
%
Our evaluation dataset derives from a largest open-source smart contract dataset~\cite{xblockcontract} as of Apirl 2024, totaling 963,151 smart contracts. 
From this original dataset, we randomly sampled 500 functions to construct the evaluation dataset. Ultimately, we filter out the functions the decompiler cannot handle, and obtain 456 pairs of source code and decompiler outputs of functions, for forming the final evaluation dataset. \textit{Noting that evaluation dataset size is similar to those used in SOTA studies~\cite{hu2024degpt,ma2024combining}}.
Here, we utilize function pairs instead of entire contracts because GPT integrated by \system{} requires shorter code snippets as input.  

Furthermore, we established the ground truth for the evaluation dataset. 
Specifically, the ground truth for function boundaries and variable types was determined by directly comparing the decompiled code with the source code, as these aspects are explicitly stated in the source code. In contrast, the ground truth for contract attributes was established through manual summarization by our domain researchers. 
To mitigate bias, the manual analysis procedure was conducted in a rigorous manner. The process consists of three steps. In the first step, six researchers with at least two years of relevant domain experience were invited and organized into three pairs. Among these, two pairs were designated as annotators, while the remaining pair, composed of researchers with at least four years of experience, served as referee experts.
In the second step, each annotator within a pair independently conducts the manual investigation. 
The third step is cross-validation and quality control. For each pair of annotators, they perform the cross-validation and discuss the labels until they obtain a consensus on the classification results.  
If consensus cannot be reached, the referee experts are consulted to provide the final decision.
In addition, we calculate the average Cohen’s Kappa coefficient to measure the consistency of labeling results, which is 0.772, thus indicating a substantial agreement between each pair of researchers.

As mentioned earlier, contract attributes refer to the semantic meaning represented by state variables (e.g., asset, address).  
To ensure comprehensive coverage of contract attribute categories, we also randomly sampled an additional 1,000 smart contracts from the original dataset and used a Large Language Model (LLM) to analyze the usage patterns of state variables within them. With the usage patterns, our experts manually summarize the representative categories as [\textit{Limit, Fee, Flag, Address, Asset, Router, Others}].

\begin{table}[h]
\scriptsize
\centering
\caption{Distribution and characteristic of the evaluation dataset.}
\setlength{\tabcolsep}{1.5mm}{
\begin{tabular}{l l}
\hline
Metric                & Evaluation dataset \\ \hline
Average lines &  163                  \\ \hline
Functionalities       &  \begin{tabular}[l]{@{}l@{}}Cross-Chain interoperability, Market Mechanisms \& Price Discovery, \\  Governance,  Risk Management \& Liquidation,  Security \& Consensus, \\ Token Management,  Value Transfer \& Incentive Distribution, \\ Entertainment \& Probabilistic Gamebling\end{tabular}                 \\ \hline
Solidity      &  From 0.4.22 to 0.8.25       \\ \hline
Deployment      &  From April 2018 to April 2024                  \\ \hline
\end{tabular} }
\label{table.detail}
\end{table}




\para{Evaluation metrics}
We utilize four metrics for evaluation:

\begin{itemize}
    
    \item \textit{function boundary match.}  We confirm the boundary prediction is correct for a given function, by manually investigating whether the starting and ending points are totally match those of sourcecode-level function.

    \item \textit{Type match.} Similar to DIRTY\cite{chen2022augmenting}, We consider a type prediction to be correct only if the predicted type fully matches the ground truth type, including data layout, and the type and name of any fields if applicable. 

    \item \textit{Contract attribute match.} Similar to type match, we determine a attribute prediction to be correct only if the predicted attribute fully matches the ground truth attribute. 

    \item \textit{Recompilation failures.} It represents errors yielded by the compiler when recompiling decompiled outputs. This indicates bugs or limitation in decompiler output~\cite{williams2020egalito}.
    
\end{itemize}


\begin{table*}[t]
\footnotesize
\caption{Comparison results between the optimized code
generated by \system{} and original decompiler output produced by the Dedaub decompiler over the evaluation dataset, for evaluating the overall effecctiveness of \system{}.}
\centering
\setlength{\tabcolsep}{1.4mm}{
\begin{tabular}{c|ccccc|ccccc|ccccc|c}
\hline
                                                              Metrics & \multicolumn{5}{c|}{Function boundary}  & \multicolumn{5}{c|}{Variable type}   & \multicolumn{5}{c|}{Contract attribute} & \begin{tabular}[c]{@{}c@{}}Recompilation failure\end{tabular} \\ \hline
\multicolumn{1}{l|}{}                                          & TP  & FP  & FN  & Prec.   & Recall  & TP   & FP  & FN  & Prec.   & Recall  & TP   & FP   & FN  & Prec.    & Recall   & Rate                                                             \\ \hline
\begin{tabular}[c]{@{}c@{}}Original \\ decompiler \end{tabular} & 316 & 149 & 310 & 67.96\% & 50.48\% & 676  & 410 & 975 & 62.25\% & 42.22\% & \multicolumn{5}{c|}{N/A}                & 100\%        \\  \hline
\system{}                                                         & 504 & 67  & 122 & 88.26\% & 80.51\% & 1349 & 113 & 252 & 92.27\% & 84.26\% & 752  & 350  & 75  & 68.06\%  & 90.93\%  & 39.78\%      \\ \hline
\end{tabular} }

{“N/A” : the corresponding tool does not support optimizing this deficiency.}

\label{table.overall}
\end{table*}



\subsection{Overall Effectiveness}
\label{sec:OverallEffectiveness}

To address RQ1, we evaluate the overall effectiveness of \system{} by comparing the optimized code generated by \system{} with the original decompiler output produced by the Dedaub decompiler (i.e., the web version of Gigahorse~\cite{grech2019Gigahorse}) using four key metrics. To achieve this, we run \system{} on the evaluation dataset. 
Additionally, we conduct a detailed analysis of the reasons behind optimization failures.
As can be seen in Table~\ref{table.overall}, the optimized code generated by \system{} significantly improves all four metrics compared to the original decompiler output. For example, the precision and recall of function boundary identification are enhanced by 20.30\% and 30.03\%, and the precision and recall of variable type inference are increased by 30.02\% and 42.04\%. 
Moreover, while the original decompiler output lacks any contract attribute information, \system{} successfully recovers the contract attributes with a precision of 68.06\% and a recall of 90.93\%. Additionally, unlike the original decompiler output, which is not suitable for recompilation, 60.22\% of the optimized code generated by \system{} can be recompiled using a compiler.


\para{Optimization failures} 
We manually inspect all the optimization failures (i.e., false positives and false negatives) of the four metrics. Firstly, we leverage the open card sorting approach similar to prior research for the error taxonomy construction. We found that SmartHalo struggles when encountering the following types of functions or contracts: E1. low-level and delegated calls, E2. inline assembly, E3. off-chain reliance, E4. inheritance structure or E5. others. Then, we compute the rates for each type of optimization failure in the above-stated taxonomy. Lastly, we conduct the root-cause mapping between the SmartHalo’s optimization failure and core components (i.e., static analysis and LLM). We found that most of optimization failures in E1 are caused by the incorrect facts generated by Gigahorse (i.e., the decompiler of SmartHalo’s static analysis module). These optimization failures can be further eliminated by integrating a more accurate decompiler in the future. We found that most of optimization failures in E2 and E5 are caused by the limited model capability of GPT-3.5. As proved by our experiment in Table~\ref{table.studyLLMs}, SmartHalo equipped with GPT-4o mini can achieve a precision rate of 91.32\% and a recall rate of 87.38\% for function boundaries, a precision rate of 90.40\% and a recall rate of 88.82\% for variable types, and a precision rate of 80.66\% and a recall rate of 91.78\% for contract attributes. Any static analysis approach like ours is fundamentally limited to addressing optimization failures in E3 and E4, as optimization failures in E3 require analyzing the off-chain data source and optimization failures in E5 require analyzing the inheritance relationship which is actually missed at the bytecode level in smart contract.






\begin{table*}[b]
\footnotesize
\caption{Comparison results between \system{}, VarLifter and SmartDagger in terms of contract attribute or variable type.}
\centering

\begin{tabular}{l|ccccc|ccccc}
\hline
Metrics     & \multicolumn{5}{c|}{Contract attribute} & \multicolumn{5}{c}{Variable type}    \\ \hline
            & TP   & FP   & FN   & Prec.    & Recall  & TP   & FP  & FN   & Prec.   & Recall  \\ \hline
SmartDagger~\cite{liao2022SmartDagger} & 111  & 364  & 641  & 23.37\%  & 10.07\% & \multicolumn{5}{c}{N/A}               \\
VarLifter~\cite{li2024varlifter}   & \multicolumn{5}{c|}{N/A}                & 115  & 31  & 1486 & 78.76\% & 7.18\%    \\ \hline

\system{}      & 752  & 350  & 75   & 68.06\%  & 90.93\% & 1349 & 113 & 252  & 92.27\% & 84.26\%                             \\ \hline
\end{tabular}
{\\ “N/A” means the corresponding tool does not support this optimization metric.}

\label{table.compare}
\end{table*}

\begin{table*}[t]
\footnotesize
\caption{Comparison results between \system{} and LLM-facilitated approach.}
\centering
\begin{tabular}{c|cllll|cllll}
\hline
 Metrics                                      & \multicolumn{5}{c|}{Function boundary}                     & \multicolumn{5}{c}{Variable type}                       \\ \hline
\multicolumn{1}{l|}{}                 & \multicolumn{1}{l}{TP} & FP  & FN  & Prec.   & Recall  & \multicolumn{1}{l}{TP} & FP  & FN   & Prec.   & Recall  \\ \hline
Original decompiler output (Baseline) & 316                    & 149 & 310 & 67.96\% & 50.48\% & 676                    & 410 & 925  & 62.25\% & 42.22\% \\
LLM without static inference prompt   & 321                    & 144 & 305 & 69.03\% & 51.28\% & 375                    & 144 & 1226 & 72.26\% & 23.42\% \\ \hline
SmartHalo                                & 504                    & 67  & 122 & 88.26\% & 80.51\% & 1349                   & 113 & 252  & 92.27\% & 84.26\% \\ \hline
\end{tabular}



\label{table.ablation}
\end{table*}

\subsection{Comparison to Prior Work}
\label{ComparisontoPrior Work}


To address RQ2, we compare \system{} with recent works. 


SigRec is not open-sourced, and we are unable to reproduce SigRec because its core algorithm (i.e., TASE) lacks disclosure of essential implementation details.
The model dataset for DeepInfer is unavailable for download; moreover, the method omits critical information regarding model inputs, architecture, hyperparameters, and training protocols. Consequently, we could not reproduce DeepInfer. Finally, we were unable to conduct a comparative evaluation between SmartHalo and Neural‑FEBI, as the publicly released Neural‑FEBI repository omits key modules and contains erroneous model artifacts.

Indeed, SmartHalo is expected to outperforms SigRec, DeepInfer and Neural‑FEBI, as they can recover only a subset of variable types/function boundaries. Specifically, Neural‑FEBI does not support boundary recovery for complex functions such as modifier functions or functions inherited from other contracts/subcontracts. Conversely, SmartHalo optimizes boundaries for all functions. And SigRec and DeepInfer only partially recover known function signatures and parameter types. Conversely, SmartHalo optimizes types for all variables

Fortunately, we contacted the authors of SmartDagger~\cite{liao2022SmartDagger} and obtained the artifact. Further, we download the artifact of VarLifter~\cite{li2024varlifter} from the open-source repository.
Hence, we run \system{} and SmartDagger over the evaluation dataset to compare their precision and recall in terms of contract attribute. And we run \system{} and Varlifter over the evaluation dataset to compare their precision and recall in terms of variable type.

As can be seen in Table~\ref{table.compare}, \system{} significantly outperforms SmartDagger and Varlifter. In terms of contract attribute, compared to SmartDagger, the precision and recall of \system{} are improved by 44.69\% and 80.86\%. Further, we manually investigate the optimization failures generated by SmartDagger. Our findings indicate that the performance of SmartDagger degrades when encountering our dataset, which includes many newly-emerging smart contracts, due to limitations in its model capability and training dataset. 
In terms of variable type, compared to VarLifter, the precision and recall of \system{} are improved by 13.51\% and 77.08\%. Similarly, we also manually inspect the optimization failures generated by VarLifter. Our investigation results shows that VarLifter can only recover a small subset of variable types, owing to its low-coverage heuristic rules. And for most variable types, the output is marked as unknown, resulting in a large number of false negatives. 

Conversely, \system{} equipped with GPT-3.5 achieves a higher precision of 68.06\% and a higher recall of 90.93\% for contract attributes, and exhibits a higher precision of 92.27\% and a higher recall of 84.26\% for variable types.  
When equipped with GPT-4o mini, \system{} achieves a even higher precision of 80.65\% for contract attributes, and presents a more higher recall of 88.81\% for variable types.




\subsection{Ablation Study}
\label{Evaluationof IndividualComponents}

Actually, \system{} is composed of two components including static analysis and large language models. To answer RQ3, we compare the Dedaub decompiler, LLM (GPT-3.5) and \system{} in terms of function boundaries and variable types, to evaluate the contribution of two components to the overall effectiveness of \system{}. Note that we do not compare them in terms of contract attributes because LLM is incapable of outputting the contract attribute without our inference prompt (e.g., contract attribute candidate).  To conduct this comparison, we further ran the LLM over the evaluation dataset. 

As can be seen in Table~\ref{table.ablation}, compared with Dedaub decompiler, LLM presents a limited improvement in precision and recall for function boundary identification. While LLM improves the precision of variable type inference by 10.01\%, but LLM drop the recall of variable type inference by 18.80\%. 
We manually inspect the optimization failures of LLM. Our manual investigation results shows that, (1) for function boundary identification, although LLM is capable of recovering a part of inherited functions, it also produces incorrect function boundary prediction which undermines the correct function boundaries in the original decompiler output due to the absence of static knowledge prompts. (2) for variable type inference, the recall of \system{} decreases rapidly, because a certain number of state variables and their types are unexpectedly deleted in the optimized code generated by LLM alone.

\begin{table*}[b]
\footnotesize
\caption{Comparison results over complex contract dataset, for evaluating \system{}'s feasibility to real-world contracts.}
\centering
\begin{tabular}{c|ccccc|ccccc|ccccc}
\hline
                                                              Metrics & \multicolumn{5}{c|}{Function boundary}  & \multicolumn{5}{c|}{Variable type}   & \multicolumn{5}{c}{Contract attribute}  \\ \hline
\multicolumn{1}{l|}{}                                          & TP  & FP  & FN  & Prec.   & Recall  & TP   & FP  & FN  & Prec.   & Recall  & TP   & FP   & FN  & Prec.    & Recall                                                                \\ \hline
\begin{tabular}[c]{@{}c@{}}Original decompiler \end{tabular} &519     & 223    &381     & 69.95\%        &  57.67\%       & 2955     & 867    & 874    & 77.31\%        & 76.18\%        & \multicolumn{5}{c}{N/A}                              \\ \hline
\system{}                                                    & 703    & 267    & 197    &  72.47\%       & 78.11\%        & 3607     & 222    & 272    & 94.20\%        & 92.98\%        & 433     & 65     & 464    & 86.95\%        & 48.27\%                        \\ \hline

\end{tabular} 

{“N/A” : the corresponding tool does not support optimizing this deficiency.}

\label{table.complex}
\end{table*}

Benefiting from the inference prompts generated by static analysis, \system{} improves the precision and recall of function boundaries by 19.23\% and 29.23\%, as well as enhances the precision and recall of variable type by 15.01\% and 60.84\%, compared to LLM alone. In conclusion, static analysis and the large language model collaboratively contribute to the improvement of decompiler output optimization in \system{}.


\subsection{Generalizability Ability}
\label{Generalizability}
To answer RQ4, we evaluate the generalizability of \system{} in terms of three aspects : 1) real-world complex contract, 2) cross-LLM applicability and 3) cross-decompiler adaptability.

\para{Feasibility to complex and diverse contracts}
As discussed in Sections~\ref{sec:DependencyExtraction} and~\ref{sec:OutputGeneration}, SmartHalo has two advantages: 1) both its SA and the LLM components are designed to be general, which enable SmartHalo to address diverse and complex real-world scenarios effectively; and (2) the prompts generated by static analysis embed rich code-context information which captures all the contents relevant to the optimization targets across the whole contract, which enables SmartHalo to perform contract-level optimization.

To evaluate this feasibility, we utilize the complex-contract dataset introduced by previous research~\cite{zheng2024dappscan}, which aggregates 682 real-world DApps from 1,199 audit reports. From this corpus, we randomly sampled 50 smart contracts (i.e., around 900 functions in total) for evaluation. Manual investigation of these 50 contracts confirmed that (1) they are highly complex, which extensively cover inheritance, modifiers, and complex storage patterns; and (2) they are sufficiently diverse, which possess more than 12 mainstream smart-contract scenarios.

As shown in Table~\ref{table.complex}, SmartHalo exhibits good performance on all evaluation metrics when optimizing these complex and diverse contracts at whole-contract optimization granularity. For example, SmartHalo improves the recall of function boundary identification by 20.44\%; increases the precision and recall of variable type inference by 16.89\% and 16.80\%; and achieve the recovery of contract attributes with a precision of 86.95\% and a recall of 48.27\%, whereas the original decompiled outputs lacks the contract attributes. 
In summary, SmartHalo can be effectively applied to a wide range of complex real-world contract types, and it exhibits good scalability to enables contract-level decompilation optimization.


\begin{table}[h]
\footnotesize
\caption{The LLMs evaluated in our study.}
\centering
\begin{tabular}{l|c|ccc}
\hline
Model Type                   & Model Name    & Open-source & Reasoning & Size \\ \hline
\multirow{4}{*}{General LLM} & GPT-3.5       & \ding{55}            & \ding{55}          & -      \\
                             & GPT-4o mini        & \ding{55}            & \ding{55}          & -      \\
                             & Llama-3 (Local) & \ding{51}            & \ding{55}          & 7B   \\
                             & Deepseek-v3      & \ding{51}            & \ding{51}          & 671B     \\ \hline
Code LLM                     & Qwen-2.5-coder         & \ding{51}            & \ding{55}          & 32B     \\ \hline
\end{tabular}
\label{table.studyLLMs}
\end{table}

\para{Applicability to different LLMs}
Here, we further investigate \system{}'s generalizability on integrating different LLMs. The study of the kind of LLM that can better drive \system{} is not the focus of our work. Conversely, we aim to explore whether \system{} maintains a good performance across different LLMs. 
To this end, we selected a diverse set of LLMs for evaluation, which considers dimensions such as general LLM versus code LLM, open-source versus closed-source models, reasoning-enabled versus non-reasoning models, and different versions of the same LLM (see Table~\ref{table.studyLLMs}). Among these LLMs, “Llama 3” is actually a small-scale model (7B parameters) that we deploy in local environment.

As shown in Table~\ref{table.diffLLMs}, SmartHalo maintains good performance across the three metrics when applying to different LLMs. 
For GPT-4o mini, the precision and recall of function boundary identification are improved by 23.26\% and 36.90\%, the precision and recall of variable type inference are increased by 28.15\% and 46.60\%, and the contract attributes are successfully identified with a precision
of 83.66\% and a recall of 90.93\%.  
For Deepseek-R1, the recall of function boundary identification is enhanced
by 41.37\%, the precision and recall of variable
type inference are increased by 19.47\% and 28.42\%, and the
contract attributes are successfully recovered with a precision
of 83.91\% and a recall of 86.11\%. 
For Qwen-2.5-coder, the precision and recall of function boundary identification are improved by 95.60\% and 85.90\%, the precision and recall of variable type inference are increased by 23.89\% and 13.31\%, and the contract attributes are successfully identified with a precision
of 70.54\% and a recall of 88.88\%. 
In particular, SmartHalo continues to perform well on the locally-deployed 7B-parameter Llama-3 model ( see the third row of Table~\ref{table.diffLLMs}), which indicates that SmartHalo can support nearly zero-cost local deployment.

In summary, SmartHalo exhibits good generalizability for a wide range of LLMs regardless of their model type, open-source status, reasoning capability, or model version.


\begin{table*}[b]
\footnotesize
\caption{Evaluation results across \system{}'s versions with different LLMs, for evaluating the \system{}'s applicability to different LLMs.}
\centering
\setlength{\tabcolsep}{1.5mm}{
\begin{tabular}{c|ccccc|ccccc|ccccc}
\hline
Setup               & \multicolumn{5}{c|}{Function boundary} & \multicolumn{5}{c|}{Variable type} & \multicolumn{5}{c}{Contract attribute} \\ \hline
\multicolumn{1}{l|}{} & TP   & FP & FN & Prec.   & Recall  & TP   & FP & FN & prec.   & recall  & TP   & FP  & FN  & prec.    & recall   \\ \hline
\system{} with GPT-3.5   & 504   &67  & 122 & 88.26\% & 80.51\% & 1349  & 113 & 252 & 92.27\% & 84.26\% & 752   & 350  & 75  & 68.06\%  & 90.93\%  \\ 
\system{} with GPT-4o mini    & 547  & 52 & 79 & 91.32\% & 87.38\% & 1422  & 151 & 179 & 90.40\% & 88.82\% & 759  & 182  & 68   & 80.66\%  & 91.78\%  \\ 
\system{} with Llama     &345      & 177   & 281   &  66.09\%       & 55.11\%        & 777     &468    & 824    & 62.41\%        & 48.53\%         & 499     & 310    & 328    & 61.68\%         & 60.34\%         \\ 
\system{} with Deepseek-R1     &575      & 397     &51    & 59.16\%        & 91.85\%        & 1131     & 253   & 470   & 81.72\%        &  70.64\%       & 626     & 120    & 201    & 83.91\%         & 86.11\%         \\ 
\system{} with Qwen &544    & 25   &82    & 95.60\%        & 86.90\%         & 889     & 143    & 712   & 86.14\%        & 55.53\%        & 735     & 307    & 92    & 70.54\%         & 88.88\%         \\ \hline
\end{tabular}
}

\label{table.diffLLMs}
\end{table*}

\begin{table*}[b]
\footnotesize
\caption{Evaluation results across different decompilers, for evaluating \system{}'s adaptability to to different decompilers.}
\centering
\setlength{\tabcolsep}{1.5mm}{
\begin{tabular}{c|ccccc|ccccc|ccccc}
\hline
                                                              Metrics & \multicolumn{5}{c|}{Function boundary}  & \multicolumn{5}{c|}{Variable type}   & \multicolumn{5}{c}{Contract attribute}  \\ \hline
\multicolumn{1}{l|}{}                                          & TP  & FP  & FN  & prec.   & recall  & TP   & FP  & FN  & prec.   & recall  & TP   & FP   & FN  & prec.    & recall                                                                \\ \hline
\begin{tabular}[c]{@{}c@{}} Dedaub \end{tabular} & 316 & 149 & 310 & 67.96\% & 50.48\% & 676  & 410 & 975 & 62.25\% & 42.22\% & \multicolumn{5}{c}{N/A}                              \\ 
\system{} on Dedaub                                                           & 504 & 67  & 122 & 88.26\% & 80.51\% & 1349 & 113 & 252 & 92.27\% & 84.26\% & 752  & 350  & 75  & 68.06\%  & 90.93\%               \\ \hline

\begin{tabular}[c]{@{}c@{}} Heimdall \end{tabular} & 316 & 149 & 310 & 67.96\% & 50.48\%         &984      &1082     &617     & 47.63\%        & 61.46\%        & \multicolumn{5}{c}{N/A}                              \\ 
\system{} on Heimdall                                                   & 440    & 109    & 186    & 80.15\%        & 70.28\%        & 1502     &834     & 99    & 64.30\%        & 93.82\%        &823      &942      & 4    & 46.63\%         & 99.52\%                       \\ \hline

\begin{tabular}[c]{@{}c@{}} Panoramix \end{tabular} &225     & 149    &401     & 60.16\%        &35.94\%         &583      &348     & 1018    & 62.62\%        &36.41\%         & \multicolumn{5}{c}{N/A}                              \\ 
\system{} on Panoramix                                                   &368     &88     &258     & 80.70\%       & 58.78\%         &1281      & 162    & 320    & 88.77\%        & 80.01\%        & 677     & 151     &150     & 81.76\%         & 81.86\%                       \\ \hline
\end{tabular} }

{“N/A” : the corresponding tool does not support optimizing this deficiency.}

\label{table.diffcompiler}
\end{table*}
\para{Adaptability to different decompilers} 
Lastly, we evaluate SmartHalo’s ability to deal with discrepancies among different Solidity decompilers. To this end, we selected three mainstream decompilers, Gigahorse (i.e., Elipmoc), Heimdall, and Panoramix, all of which have been widely applied in prior studies and industry~\cite{grech2022Elipmoc,su2025disco}. Specifically, we run these three decompilers on the 456 functions contained in our evaluation dataset to obtain their respective decompiled outputs. Subsequently, SmartHalo is run on the decompiled code produced by each decompiler, and the results of three evaluation metrics are collected for comparative analysis.

As illustrated in Table~\ref{table.diffcompiler}, evaluation results show that SmartHalo improves both precision and recall across all the metrics for different decompilers. For Heimdall, the precision and recall of function boundary identification are enhanced by 12.19\% and 19.80\%, the precision and recall of variable type inference are increased by 16.67\% and 32.36\%, and the contract
attributes are successfully recovered with a precision of 46.63\% and a recall of 99.52\%, whereas the original decompiled outputs lack the contract attributes. 
For Panoramix, the precision and recall of function boundary identification are improved by 20.54\% and 22.84\%, the precision and recall of variable type inference are increased by 26.15\% and 43.60\%, and the contract
attributes are successfully identified with a precision of 81.76\% and a recall of 81.86\%, whereas the original decompiled outputs lack the contract attributes.
These results indicate that SmartHalo is generic to different Solidity decompilers.

\subsection{Efficiency}
\label{Efficiency}

To address RQ5, we execute SmartHalo on our evaluation dataset and measure both its runtime and monetary costs.

As shown in Table~\ref{table.efficiency}, the proposed approach can establishes high-quality optimization with an efficient and cost-effective manner. For example, we compute the average time for
each step of \system{}: 1.94 s for the dependency-based semantic extraction module, 8.95 s for the LLM-driven semantic enrichment module and 13.10 s for the correctness verification module.
%
%
For monetary costs, the GPT-4o mini-powered version of SmartHalo consumes an average of 8,920 tokens for each function, which incurs an average expense of \$0.00136. 
With our statistical analysis, the average number of functions per contract within the evaluation dataset is approximately 34. Accordingly, we can roughly estimate that, \system{} costs an average expense of \$0.046 and an average runtime of 815.66 s per smart contract.

In addition, with the rapid development of LLMs, an increasing number of open-source alternative LLMs has emerged. As mentioned earlier, SmartHalo can be deployed with a local, cost-free, small-scale LLama-7B model and achieve good performance, as shown in Table~\ref{table.diffLLMs}. In this way, SmartHalo’s monetary cost can be reduced to nearly zero.

In sum, \system{} is proved to be cost-effective in terms of runtime and monetary costs, which indicate its practical value.

\para{Parameter setting influence} Further, we evaluate how the settings of iteration limit influence the cost of correctness verification. Specifically, we run the correctness verification module with different iteration limit ($n = 1,2,3…$), and compute the convergence rate, runtime and monetary costs of evaluated functions. The evaluation results show that, when $n \ge 3$, both the runtime and monetary costs stop increasing, as the convergence rate of correctness verification module reaches 100\%. Hence, we set the iteration limit $n$ to 3 in our evaluation setup.

\begin{table}[t]
\footnotesize
\caption{The average time and tokens for analyzing each method in the evaluation dataset.}
\centering
\begin{tabular}{lcc}
\hline
Efficiency of \system{} & Avg.time(s) & Avg.tokens \\ \hline
Dependency-based semantic extraction     &1.94             & -           \\ 
LLM-driven semantic enrichment     &8.95             & 4272 (\$ 0.00065)           \\ 
Correctness verification     & 13.10            & 4648 (\$ 0.00071)          \\ \hline
Total     & 23.99            & 8920 (\$ 0.00136)          \\ \hline
\end{tabular}
\label{table.efficiency}
\end{table}

\begin{figure*}[b]
\begin{minipage}{\textwidth}
\vspace{0.15in}
 \begin{minipage}[t]{0.47\textwidth}
  \centering
  \footnotesize
     \makeatletter\def\@captype{table}\makeatother\caption{Comparison results for evaluating the effectiveness of \system{} on the Reentrancy detection}
\begin{tabular}{c|ccc}
\hline
Reentrancy                  & \multicolumn{3}{c}{Precision} \\ \hline
                  & TP      & FP     & rate       \\ \hline
SliSE         &70      &27      & 72.16\%    \\
SliSE+\system{} &78      &19      & 80.41\%    \\ \hline
\end{tabular}
\label{table.downstreamforcrosschain}
  \end{minipage} \hspace{0.1in}
  \begin{minipage}[t]{0.47\textwidth}
   \centering
   \footnotesize
        \makeatletter\def\@captype{table}\makeatother\caption{Comparison results for evaluating the effectiveness of \system{} on the attack identification.}
\begin{tabular}{c|ccc}
\hline
Attack contract identification                  & \multicolumn{3}{c}{Recall} \\ \hline
                  & TP      & FN     & rate       \\ \hline
BlockWatchdog         & 15     & 3     & 83.33\%    \\
BlockWatchdog+ \system{} & 18     & 0     & 100.00\%    \\ \hline
\end{tabular}
\label{table.downstreamforcodereuse}
   \end{minipage}
\end{minipage}
\end{figure*}
\begin{table*}[b]
\footnotesize
\caption{Comparison results for evaluating the effectiveness of \system{} on Integer-overflow vulnerability.}
\centering
\begin{tabular}{c|ccc|ccc}
\hline
Integer overflow vulnerability                  & \multicolumn{3}{c|}{Precision}   & \multicolumn{3}{c}{Recall} \\ \hline
                  & TP      & FP     & rate    & TP      & FN     & rate   \\ \hline
Mythril            &13      &5       & 72.22\%   &13      &37       & 26.00\%  \\
Mythril+ \system{} &32      &2       & 94.18\%   &32      &18       & 64.00\%  \\ \hline
\end{tabular}


\label{table.overflow}
\end{table*}

\section{Effectiveness for Downstream Task}
\label{Effect of Data}


These experiments were designed to assess how  the contract attributes, function boundaries, and variable types recovered by \system{} contribute to their respective downstream tasks.

\begin{figure*}[t]

\includegraphics[width=7.2in]{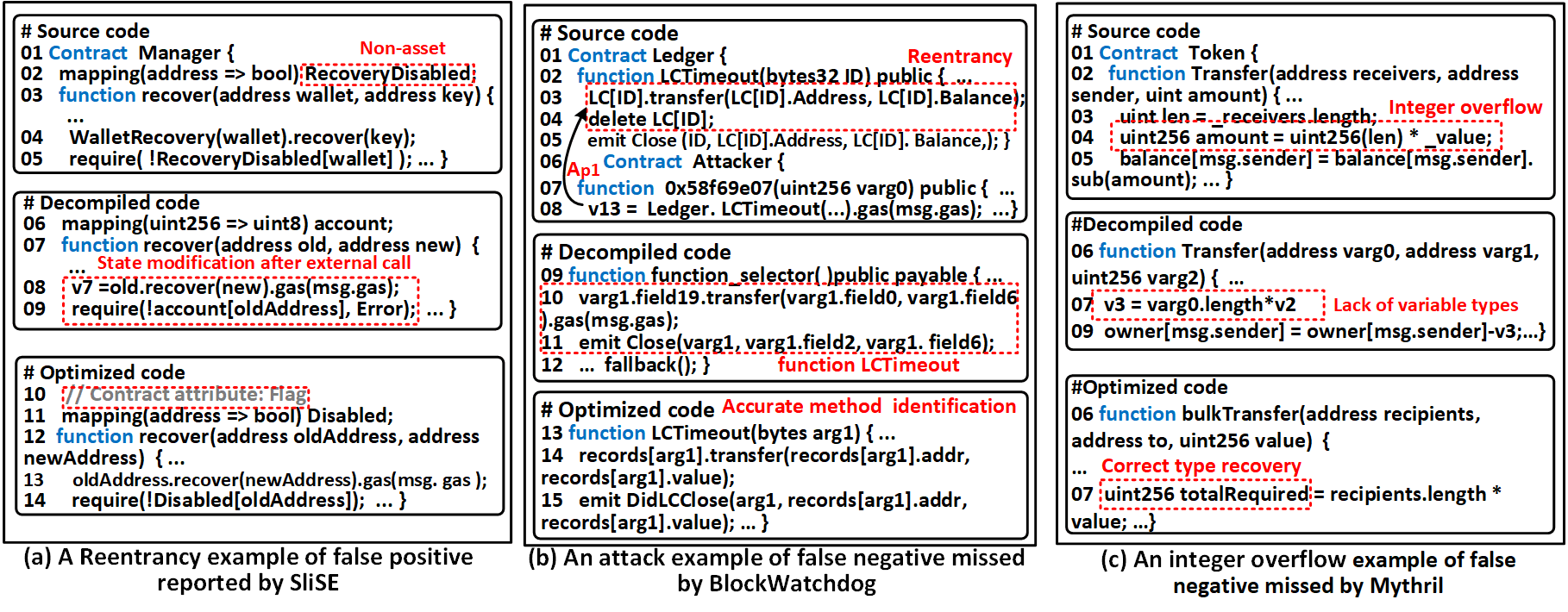}
\caption{The sample false alarms reported by SOTA tools. These FPs and FNs can be eliminated with the integration of \system{}. }
\label{falsealarm}
\vspace{-0.1in}
\end{figure*}

\para{Contract attributes for Reentrancy vulnerability detection}
Prior research~\cite{liao2022SmartDagger} shows that the effectiveness of contract attribute recovery reflects on eliminating the false positives for Reentrancy detection.
Hence, We evaluate how the contract attributes optimized by \system{} contribute to improving the precision of Reentrancy detection. To this end, 
we utilize the latest manually-labeled DApp dataset introduced in previous research~\cite{wang2024efficiently}, with a total of 81 positive labels for Reentrancy vulnerabilities. 
To conduct this evaluation, we run the SOTA tool SliSE~\cite{wang2024efficiently} and another tool that integrates SliSE with \system{} over the manually-labeled DApp dataset to evaluate their precision. 
As can be seen in Table~\ref{table.downstreamforcrosschain}, SliSE+\system{} improve the precision to 80.41\%, whereas the precision of SliSE is only 72.16\%. We further manually inspect all the false positives reported by these two tools. We found that 8 of 27 false positives reported by SliSE can be eliminated by \system{}, with the help of more accurate contract attributes recovered by SliSE+\system{}.
Fig.~\ref{falsealarm}(a) illustrates a Reentrancy example of a false positive reported by SliSE. In this instance, SliSE erroneously flags the  function \textit{recover} due to its adherence to a pre-defined Reentrancy pattern, where function \textit{recover} modifies state variables after the external call (lines 8 and 9). Upon meticulous inspection of the source code, we found that the contract attribute of the state variable \textit{RecoveryDisable} belong to \textit{[non-asset]} (line 02). This attribute indicates that the \textit{recover} cannot be exploited for profit gain by adversaries, thus \textit{recover} does not pose a Reentrancy risk. In contrast, SmartHalo accurately identifies the contract attribute of this state variable as [Flag] (line 10), effectively eliminating this false positive.



\para{function identification for attack identification}
Prior research~\cite{yang2024uncover} shows that  attack identification relies heavily on the precise cross-contract call flow analysis between victim and attacker contracts, which effectively models the attack path. Incorrect function identification can lead to missed external calls and corresponding call flows, thereby impeding the identification of attack path. Therefore, the effectiveness of function boundary optimization is reflected in reducing false negatives in attack identification.
To assess how function boundaries optimized by \system{} enhance the recall of attack identification, we utilize the latest attack contract datasets introduced by previous research~\cite{yang2024uncover}, comprising 18 pairs of attacker and victim contracts from 15 real-world incidents. 
Further, we compare the SOTA tool BlockWatchdog~\cite{yang2024uncover} with an enhanced version that integrates \system{} with BlockWatchdog. As shown in Table~\ref{table.downstreamforcodereuse},  BlockWatchdog+SmartHalo enhances the
recall by 16.67\% compared with BlockWatchdog alone.
Additionally, we manually inspect all the false negatives reported by BlockWatchdog, finding that BlockWatchdog+\system{} successfully eliminates all false positives through more accurate function identification optimized by \system{}. Fig.~\ref{falsealarm} (b) presents an example of false negative missed by BlockWatchdog. In this case, function \textit{LCtimeout} contains a Reentrancy vulnerability, where it logs the transfer results (line 4) after transferring the asset (line 3). Further, an adversary could exploit this by invoking the function \textit{LCTimeout} to trigger the Reentrancy vulnerability (path $A_{p1}$). BlockWatchdog fails to detect this attack path (i.e., false negative), because function \textit{LCtimeout} is unavailable for BlockWatchdog at the decompiled code level~\cite{yang2024uncover}. In contrast, \system{} accurately identify the function \textit{LCTimeout} (line 13-15), thereby eliminating this false negative.

\para{Variable types for Integer-overflow vulnerability detection}
Prior research~\cite{lai2020static, sun2021mutation} demonstrate that  detecting integer-overflow vulnerabilities heavily depends on accurately identifying the upper and lower bounds of variables based on their types. Incorrect variable type recovery results in both false positives and false negatives ~\cite{lai2020static}. Therefore, the effectiveness of variable type optimization is reflected in its ability to reduce false negatives and false positives in integer-overflow vulnerability detection.
We further evaluate the effectiveness of the variable types optimized by SmartHalo on the precision and recall of integer-overflow vulnerability detection. For this evaluation, we utilize integer overflow datasets collected from real-world attacks in previous research~\cite{zhou2020ever}, which includes 50 vulnerable contracts with integer overflow vulnerabilities.
Further, we compare the SOTA tool Mythril~\cite{mythril} with a modified version that integrates Mythril with \system{}. As shown in Table~\ref{table.downstreamforcodereuse},  Mythril+\system{} enhances the
precision by 21.96\% and the recall by 38.00\% compared with Mythril alone.
Additionally, we manually investigate all the false negatives and false positives reported by Mythril. We found that 60\% of the false positives and 51.35\% of the false negatives reported by Mythril
can be eliminated by Mythril+\system{}, thanks to the more accurate variable types optimized by \system{}. Fig.~\ref{falsealarm} shows an example of false negative missed by Mythril. 
In this case, the function \textit{batchTransfer} contains a \textit{uint256} variable \textit{amount}, which is correlated with two arguments, \textit{receivers} and \textit{value}, both of which can be passed arbitrarily via external calls (line 4), leading to an integer overflow vulnerability.

Mythril detects integer overflow based on two conditions: (1) scanning for key EVM instructions (e.g., ADD, SUB, MUL, etc.), and (2) invoking Z3 solver with the maximum bit-width derived from the variable type to determine whether integer overflow occurs. Due to the absence of type information for the variable \textit{v3} (line 07) in the decompiled code, Mythril fails to identify the maximum bit-width, resulting in a false negative report. 
Conversely, \system{} correctly identifies the type of variable \textit{v3} as a \textit{uint256} type (line 07), thus eliminating the false negative.

\section{Discussion and Limitation}
\label{Discussion}

\system{} shares the following advantages in decompiler output optimization: (1) As evidenced by evaluation, \system{} is obviously effective in optimizing function boundaries, variable types and contract attributes, demonstrating superior performance compared to previous approaches; 
(2) \system{} establishes a novel combination of static analysis (SA) and large language models (LLMs), enabling the joint optimization on the decompiler output with the advantages of both SA and LLMs, thereby overcoming the limitation of prior works.
(3) \system{} implements a rigorous correctness verification for the optimization output to eliminate the optimization errors. 



SmartHalo has good adaptability to newly-deployed contracts due to its flexible and generic core components:  (1) Static analysis employs typical control‑flow analysis to extract the dependencies that are generic to all contracts.  These dependency‑extraction methods are equally applicable to newly deployed contracts; (2) The LLMs is adept at few-shot in-context learning, which refers to the capability of LLMs to instantaneously comprehend and address new and unseen cases/tasks during the inference stage, based solely on a small number of known examples (shots). This helps SmartHalo to be feasible to the newly-deployed contracts. 

Although our evaluation of SmartHalo focuses on the function level, \system{} is actually capable of supporting contract‑level optimizations. Specifically, the prompts generated by static analysis embed complete code-context information which captures all the contents relevant to the optimization targets across the whole contract, which enables SmartHalo to perform contract-level optimization. For example, the prompt templates of Table~\ref{table.COT} introduce the contextual information of modifiers (i.e., row 10), which guides the LLM to reconstruct the corresponding modifier functions; and as highlighted by the yellow part of Fig. 6(a), the COT prompts produced by static analysis effectively capture the dependencies between key–value pairs within complex storage patterns (e.g., mappings and structs), this capability facilitates the recovery of the complex storage variable types ( i.e., mapping (bytes 32 = uint256) data). Further, as evidenced by Section~\ref{Generalizability}, we have evaluated SmartHalo’s feasibility to complex and diverse contracts. And the results confirm that SmartHalo exhibits good scalability for contract-level decompilation optimization.

The evolution of LLM APIs presents potential challenges to the reproducibility of evaluation results. To address this concern, we have implemented the following strategies. Firstly, we explicitly list the exact API configurations for each model in the SmartHalo’s repository. These details encompass the server provider, model identifier, version/release date, invocation endpoints, context length limitations, and relevant environment variables. Further, for open-source models, except for API configurations, we provide an access link of the repository corresponding to each model version used in this paper. Finally, as evidenced by Section V-E, SmartHalo maintains a good performance on the evaluated smaller language model. In future work, we can explore integrating more alternative fine-tuned smaller language models into SmartHalo to further reduce the effects of evolving LLMs on reproducibility.

A current limitation is that SmartHalo cannot recover inheritance structures within contracts. The reason for this limitation is that, inheritance relationship and class information are missed at the bytecode level in smart contracts. Particularly, state-of-the-art efforts [9], [19] in other domains (e.g., C++, Java) also cannot recover the inheritance structure for smart contract, because they mainly rely on class information. Given the difficulty of this task, we leave the exploration of inheritance structure recovery to future work.


\para{Threats to validity} 
An external threat to validity is that the number of smart contract functions used in our evaluation here is relatively small (i.e., 456 smart contract functions). 
However, we consider the dataset size sufficient for the following reasons.
Firstly, the dataset was randomly selected from the largest real-world contract repository.
Secondly, the dataset size is similar to those used in SOTA studies~\cite{hu2024degpt,ma2024combining}.
Thirdly, our manual investigation confirms that the dataset involves diverse and mainstream application scenarios of smart contract\footnote{Note that the manual investigation results are available via \url{https://figshare.com/s/5d4fc1ad2312a4be9370?file=49346911}}. 


An internal threat to validity is the predefined type rules employed by \system{} for correctness verification (Section~\ref{sec:CorrectnessVerification}).
SmartHalo remains a highly generalizable framework for the following reasons:
First, SmartHalo focuses on optimizing Solidity decompiler output. This setup is aligned with recent contract analysis research~\cite{liao2024smartaxe, liu2024using}.
Second, most contracts are written in Solidity~\cite{liao2024smartaxe} and are widely deployed on major blockchains (e.g., Ethereum, TRON, and BNB).
Third, the static rules(see Figure~\ref{fig:StaticRule}) are generic across different Solidity versions, as they reflect knowledge of fundamental types (e.g.,array). A review of Solidity’s release-changelogs indicates that these rules remained unchanged across all Solidity’s major updates~\cite{Solidity_release_changelogs}.
Fourth, SmartHalo is backward-compatible, can easily adapt to contract evolution with corresponding knowledge.

\section{Related Work}
\para{Smart Contract Decompilation} 
The decompilation of smart contracts has garnered significant attention due to the prevalence of bytecode-only smart contracts~\cite{suichePorosity}
Gigahorse~\cite{grech2019Gigahorse} translates smart contracts from low-level EVM bytecode into a 3-address code representation. Elipmoc~\cite{grech2022Elipmoc} is the 2.0 verson of Gigahorse, which employs a novel context sensitivity called transactional sensitivity to achieve a more effective static abstraction. Erays~\cite{zhou2018Erays} and EtherSolve~\cite{contro2021EtherSolve} aim to recover the CFG from the EVM bytecode. Recently, SigRec~\cite{chen2022SigRec} has been proposed to recover the signatures of public functions. DeepInfer~\cite{zhao2023DeepInfer} leverages deep learning techniques to infer function signatures. 

\para{Decompilation Optimization} 
The existing decompilation optimization research mainly focus on other languages such as C++ and Java~\cite{pei2020trex,peng2024fast}, which can be divided into the following two categories~\cite{pang2021sok}. The first category aims at recovering the semantic information from the assembly code or intermediate language. Among them, previous works, including DEBIN~\cite{he2018debin}, 
OSPREY~\cite{zhang2021osprey}, and BDA~\cite{zhang2019bda}, analyze the program dependency via the static analysis. However, they are with low coverage due to their reliance on heuristic rules. Another group of works, including Nero~\cite{david2020neural}, NFRE~\cite{gao2021lightweight}, SYMLM~\cite{jin2022symlm}, focus on designing encoder-decoder architecture models to predict the function name for the binary. 
The other category of work focuses on optimizing variable name, variable type, and structure of code generated by decompiler, including DIRE~\cite{lacomis2019dire}, DIRTY~\cite{chen2022augmenting}
and DeGPT~\cite{hu2024degpt}. 



\section{Conclusion}
\label{sec:conclusion}

In this paper, we introduced SmartHalo, a novel framework designed to address critical deficiencies in existing Solidity smart contract decompilers. By leveraging the strengths of static analysis (SA) and large language models (LLMs), SmartHalo significantly enhances the accuracy and readability of decompiled code. 
%
Extensive evaluation demonstrates that SmartHalo outperforms state-of-the-art decompilers, showing substantial improvements in function boundary identification, variable type inference, and contract attribute recovery. We also showed that the output of \system{} can significantly enhance the effectiveness of downstream tasks.





\normalem

{
    \bibliographystyle{ieeetr}
    \bibliography{bib}
}




\end{document}